\documentclass[prd,nofootinbib,superscriptaddress,twocolumn]{revtex4}

\usepackage{amsfonts,amssymb,amsthm,bbm,amsmath}
\usepackage{color,psfrag}
\usepackage[dvips]{graphicx}
\usepackage{hyperref}

\newcommand{\nocontentsline}[3]{}
\let\origcontentsline\addcontentsline
\newcommand\stoptoc{\let\addcontentsline\nocontentsline}
\newcommand\resumetoc{\let\addcontentsline\origcontentsline}

\makeatletter
\def\l@subsubsection#1#2{}
\makeatother
\def\nn{\nonumber}

\def\be{\begin{equation}}
\def\ee{\end{equation}}

\newcommand{\p}{\partial}
\newcommand{\pp}{\partial}

\def\cL{ {\cal L}}
\def\cF{ {\cal F}}
\def\cT{ {\cal T}}
\def\cHeff{ {\cal H}_{\textrm{eff}}}

\def\a{\alpha}
\def\b{\beta}

\def\D{\Delta}

\def\s{\sigma}

\def\bra{\langle}
\def\ket{\rangle}
\def\l{\left}
\def\r{\right}
\def\f{\frac}
\def\MO{\mathcal {O}}

\def\MA{\mathcal {A}}

\def\Seff{S_{\mathrm{eff}}}
\def\Heff{H_{\mathrm{eff}}}
\def\rhoeff{\rho_{\mathrm{eff}}}
\def\peff{p_{\mathrm{eff}}}

\def\rd{\mathrm{d}}
\def\newr{\mathfrak{r}}
\def\err{\ell}
\def\radius{r_s}
\def\Teff{{\mathcal T}^{{\rm eff}}}

\begin{document}
\title{Effective Dynamics of Spherically Symmetric Static Spacetime} 
\author{Etera R. Livine}
\email{etera.livine@ens-lyon.fr}
\affiliation{ENS de Lyon, CNRS, Laboratoire de Physique (LPENSL), F-69342 Lyon, France}
\affiliation{Perimeter Institute for Theoretical Physics, 31 Caroline Street North, Waterloo, ON, N2L 2Y5, Canada}
\author{Yuki Yokokura}
\email{yuki.yokokura@riken.jp}
\affiliation{RIKEN Center for Interdisciplinary Theoretical and Mathematical Sciences (iTHEMS), Wako, Saitama 351-0198, Japan}
\affiliation{ENS de Lyon, CNRS, Laboratoire de Physique (LPENSL), F-69342 Lyon, France}

%%%%%%%%%%%%%%%%%%
\begin{abstract}
In general relativity, the Einstein equations provide spherically symmetric static spacetimes with dynamics defined as an evolution along the radial coordinate $r$. The geometrical sector becomes a one-dimensional mechanical system, with the Misner-Sharp mass and lapse as canonically conjugate variables, and a vanishing Hamiltonian for pure gravity. Coupling classical or quantum matter fields, or introducing (quantum) corrections to general relativity, then generate a non-vanishing effective Hamiltonian, leading to non-trivial evolutions of the mass and lapse. We illustrate this mechanism through various examples of classical matter fields and identify Hamiltonians describing the effective dynamics of gravity coupled to perfect fluids with linear barotropic equation of state. Finally, we derive effective Hamiltonians that reproduce the gravitational semi-classical dynamics coupled to renormalized quantum matter fields and discuss the conditions for which the singularity at $r=0$ is resolved. In particular, we find a singularity-free black-hole-like solution, stabilized by quantum matter, smoothly transitioning from a bulk with constant negative Ricci scalar to the standard outside Schwarzschild metric. This opens new possibilities for the modeling of both semi-classical corrections and deep quantum effects on the interior structure of self-gravitating compact objects and black holes.

\end{abstract}

\maketitle

\setcounter{secnumdepth}{2}
\setcounter{tocdepth}{2}
\tableofcontents

%%%%%%%%%%%%%%%%%%%%%%%%%%%%%%%%%%%%%%%%%%%%%%%%%%%%%%%%%%%%%%%%%%%%%%%%%%%%%%%%%%%%

\section{Introduction}
\label{sec:intro}

Gravity universally interacts with all types of matter and fields, inevitably intertwining their dynamics in a complex non-linear dance. 
%
%and all the dynamics mixes together. This is the manifestation of universal gravitation, which gives rise to non-linear complex dynamics, makes the energy of matter and gravity inseparable, and produces, classically, singularities in a extremely strong gravity. 
%
Spherically symmetric static space-times provide the most basic arena for investigating such universal dynamics of gravity. Due to its role as universal equilibrium (non-rotating) geometries, this symmetric-reduced configuration plays, despite its simplicity, a significant role in various contexts, among which the analysis of the internal structure of self-gravitating compact objects (e.g. \cite{Tolman:1939jz,Oppenheimer:1939ne,Buchdahl:1959zz,Cardoso:2019rvt}), %the study of the properties and fate of gravitational singularities \cite{XX}, and 
the exploration of quantum black holes (e.g. \cite{Kuchar:1994zk,Ashtekar:2005qt,Haggard:2014rza}) and more generally of the quantum gravity regime (e.g. \cite{Thiemann:1992jj,Bojowald:2005cb}). In particular, consideration of a radial Hamiltonian for spherically symmetric spacetime is receiving
attention recently (e.g.\cite{Bodendorfer:2019cyv,Perez:2023jrq,BenAchour:2023dgj, MenaMarugan:2024, Koch:2025yuz}).
%
%Spherically symmetric static spacetime provides the simplest and most basic arena for investigating dynamics of gravity. In fact, it plays significant roles in many contexts: analyzing the internal structure of self-gravitating objects such as stars, studying the properties of singularities, investigating the quantum properties of black holes, and exploring quantum gravity. In the case of pure Einstein gravity, there is only one solution, the Schwarzschild solution. However, once other degrees of freedom such as matter fields or quantum gravitational effects are involved, the dynamics becomes difficult to solve. 
%
In this context, we investigate how to model the feedback of classical and quantum matter fields, as well as modified gravitational dynamics, on spherically symmetric static spacetime through simple consistent action principles and effective Hamiltonians, and how these modify the structure of black hole solutions.

%Indeed, the coupled system of gravity plus matter is described by a pair of equations: on the one hand, the equation for the matter dynamics (evolving in the geometry) and, on the other hand, the Einstein equations for the dynamics of the geometry (in presence of matter),
%\be
%\label{eqn:firsteqn}
%G_{\mu\nu}=8\pi G \mathcal{T}_{\mu\nu}\,,
%\ee
%where the tensor $\mathcal{T}_{\mu\nu}$ can be a classical energy-momentum $T_{\mu\nu}$ or a renormalized expectation value $\bra \Psi |\hat{T}_{\mu\nu}|\Psi\ket$, or even further include modified gravity corrections.
%%
%Both equations involve both matter and geometry, creating a non-linear feedback of one onto the other leading to complex dynamics. A strategy can be to first solve the matter dynamics equation, express the matter fields in terms of the geometry, and then plug this back into the Einstein equations. We would then obtain an effective equation driving the dynamics of the space-time metric accounting for the coupling of geometry with matter fields and all other relevant degrees of freedom. This is the path that we propose to develop further in the case of spherically symmetric static spacetime.

%\YY{
Indeed, the coupled system of gravity plus matter is described by a pair of equations: on the one hand, the equation for the matter dynamics (evolving in the geometry) and, on the other hand, the Einstein equations for the dynamics of the geometry (in presence of matter),
\be
\label{eqn:firsteqn}
G_{\mu\nu}[g]
\,=\,
8\pi G\, \mathcal{T}_{\mu\nu}[g,\phi]\,.
\ee
The Einstein tensor $G_{\mu\nu}$ depends solely on the metric $g$. The source tensor $\mathcal{T}_{\mu\nu}$ depends on both the metric $g$ and matter degrees of freedom summarized as $\phi$. It can be a classical energy-momentum $T_{\mu\nu}$ or a renormalized expectation value $\bra \Psi |\hat{T}_{\mu\nu}|\Psi\ket$, or further include modified gravity corrections.
Both the matter and the Einstein equations involve both matter and geometry, creating a non-linear feedback of one onto the other leading to complex dynamics. A strategy is to first solve the matter dynamics equation, express the matter fields in terms of the geometry, and then plug this back into the Einstein equations. We would then obtain an effective equation driving the dynamics of the space-time metric accounting for the coupling of geometry with matter fields and all other relevant degrees of freedom,
\be
\label{eqn:secondeqn}
G_{\mu\nu}[g]
\,=\,
8\pi G\, \Teff_{\mu\nu}[g]\,.
\ee
This modified gravity equation would involve an effective stress-energy tensor $\Teff_{\mu\nu}$. It would now depend solely on the metric, but would nonetheless have coupling constants depending on relevant parameters and charges of the matter distribution, such that the solutions of this equation would match exactly the solutions of the original Einstein equation.
This is the path that we propose to develop further in the case of spherically symmetric static spacetime.
%}

More precisely, we consider the following metric ansatz for spherically-symmetric static space-time\footnotemark{}:
\begin{equation}
\label{metric}
\rd s^2
=
-f(r)b(r)^2 \rd t^2 +f(r)^{-1}\rd r^2+r^2\rd \Omega^2,
%~~f(r)=1-\frac{a(r)}{r}.
\end{equation}
where the metric components are solely functions of the radial coordinate $r$.
\footnotetext{In other works (e.g. \cite{Bodendorfer:2019cyv,BenAchour:2023dgj,MenaMarugan:2024}), a different form of the line element ansatz is used: 
%A slight change of gauge is the line element ansatz:
\be
\rd s^2
=
-F(R)\rd t^2 +F(R)^{-1}c(R)^2\rd R^2+r(R)^2\rd \Omega^2.
\nn
\ee
Here, addition to the lapse $c(R)$ and a function $F(R)$, $r(R)$ is a function of a new radial coordinate $R$. The number of degrees of freedom seems different from ours, but after fixing the gauge, they are the same. Indeed, setting $c(R)=1$ is a gauge fixing, and the correspondence is given by $\rd_R r=b(r)$ and $F(R)=f(r(R))b(r(R))^2$. Note that, at least, one of the key results, \eqref{Sg}, holds even in a full covariant formalism \cite{PY}.}
%where $F$ and $r$ are now functions of the new radial coordinate $R$. The correspondance is highly non-linear and explicitly given by $\rd_R r=b(r)$ and $F(R)=f(r(R))b(r(R))^2$.}
%
The field $b(r)$ is the lapse up to the factor $f(r)$. It is more exactly the area element of the 2D surface normal to the spatial sphere with radius $r$.
Further introducing the field $a(r)$ such that $f(r)=1-a(r)/{r}$,
%or equivalently $a(r)=r(1-f(r))$,
we recognize $m(r)=a(r)/2G$ as the Misner-Sharp mass contained in the ball of radius $r$. It is a locally conserved energy (even in dynamical cases \cite{Kodama}) with contributions from intrinsic mass and gravitational energy, which agrees with the ADM energy in the $r\to \infty$ limit for asymptotically flat spacetimes \cite{MTW,Hayward:1994bu}. 

The reduced Einstein-Hilbert action evaluated on this metric ansatz takes a simple form, as will be shown: 
\begin{equation}
\label{Sg}
    S_{\mathrm{g}}[a,b]=\frac{1}{2G}\int^{r_2}_{r_1} dr b(r) \dot a(r),
    \quad\textrm{with}\,\,
    \dot a\equiv \frac{\rd a}{\rd r},
\end{equation}
where we dropped a time interval factor and a surface term.
%We use the notation $\dot x\equiv {\rd x}/{\rd r}$.
%where $a(r)\equiv 2G m(r)$, $\dot x\equiv \frac{dx}{dr}$, and the canonically conjugate variable $b(r)$ is the volume element of 2D spacetime part normal to the spatial sphere with radius $r$ (Sec.\ref{sec:gravity}).
This one-dimensional mechanical system, whose evolution along $r$ is obviously with a vanishing Hamiltonian, has trivial equations of motion, $\dot a=\dot b=0$. This leads to the Schwarzschild metric as the unique solution.
%
%Indeed, this gives the Schwarzschild metric as the solution. 

Therefore, matter fields, modified gravity corrections, or quantum fluctuations, would all generate an effective Hamiltonian $\Heff[a,b,r]$, leading to non-trivial dynamics for the canonical pair, the mass $a(r)$ and lapse factor $b(r)$.
%
%This seems like a mechanical system with a zero Hamiltonian, and therefore, deviations from the vacuum Einstein-Hilbert gravity lead to a non-zero Hamiltonian, that is, an effective Hamiltonian $\Heff(a,b,r)$.
Indeed, the effective action $\Seff$ given by 
\begin{equation}
\label{Seff0}
    S_{\mathrm{eff}}[a,b]=\int^{r_2}_{r_1} dr \left(\frac{b(r) \dot a(r)}{2G}-\Heff[a,b,r]\right),
\end{equation}
leads to the Hamilton equations of motion:
\begin{equation}\label{Heq}
\dot a = 2G \frac{\partial \Heff}{\partial b},
\qquad
\dot b = - 2G \frac{\partial \Heff}{\partial a}\,.
\end{equation}
These equations would lead back to the exact same dynamics for the geometry, as prescribed by the Einstein equations \eqref{eqn:firsteqn} coupling geometry to the considered energy-stress energy $\cT_{\mu\nu}$.

The solutions $a(r),b(r)$ to these equations then define spherically symmetric static metrics, which can be understood as modified Schwarzschild metrics for self-gravitating objects. Here, instead of cataloging such metrics directly by their components $a(r),b(r)$, we will adopt the logic of classifying models according to the effective Hamiltonian generating those metric components. In the long run, this should allow simpler links to modified gravity actions in general relativity.

We thus embark on the task of exploring the physics of such effective Hamiltonians. More precisely, we aim at understanding which effective Hamiltonians result from the coupling of geometry with various matter sources (such as a scalar field, the electromagnetic field, or more general matter fields), and at investigating how quantum field renormalization affects $\Heff$. Then, we wish to analyze the properties of spacetime for various choices of dynamics, and in particular to identify classes of Hamiltonian leading to singularity-free black hole solutions.

\medskip

First, in section \ref{sec:gravity}, we explain in detail the spherical symmetric static reduction of general relativity and compute the resulting reduced Einstein-Hilbert action with the canonical pair of mass-lapse variables $(a,b)$ and a vanishing Hamiltonian for pure gravity.
This leads us to introduce the concept of a non-vanishing effective Hamiltonian $\Heff[a,b,r]$, that would encode the non-trivial evolution $a(r),b(r)$ in terms of the radial coordinate $r$ resulting from the coupling of geometry to matter fields or quantum fluctuations or other potential sources.

Then, in section \ref{sec:classical}, we work out explicitly the effective Hamiltonian induced by the dynamical coupling of classical matter fields to  geometry, more specifically, by considering the cases of a cosmological constant, a massless scalar field, a Maxwell field, and non-linear extensions of electromagnetism. This gives substance to our proposal.
%we examine the general structure and properties of $\Heff(a,b,r)$. We start with classical examples: cosmological constant, massless scalar field, Maxwell field, and non-linear electromagnetic theory (Sec.\ref{sec:classical}). 
%
%Motivated by these results, we consider two types of $\Heff(a,b,r)$ and study their behavior (Sec.\ref{sec:general}).
%
This leads us to a more in-depth analysis of the compatibility of effective Hamiltonians with the Einstein equations for general relativity coupled to a perfect fluid with linear equation of state in section \ref{sec:general}. This allows us to sharpen the general admissible shape of our effective action principles.
%
%The dynamical equation \eqref{eom} restricts the structure of $\Heff(a,b,r)$ at some level.

\smallskip

Next, we focus on a more specific class of effective Hamiltonians,
\begin{align}
\label{Heff}
     \Heff[a,b,r]&=b \rhoeff[a,r],\\
\label{rhoeff}
     \rhoeff[a,r]&=\frac{1}{2G}
     \sum_{n\ge 0} c_n(r)a(r)^n\,,
%     \rhoeff(a,r)&=\frac{1}{2G}\left(c_0(r)+c_1(r)a(r)+\frac{1}{2}c_2(r)a(r)^2+\cdots\right),
\end{align}
such that $\Heff$ is linear in $b$. Here, $\rhoeff\equiv -4\pi r^2 \Teff{}^t{}_t$ is the 1D energy density, and the coefficient functions $c_n(r)$ characterize the system of interest.
This ansatz is consistent with the logic that the lapse factor $b$ plays the role of the Lagrange multiplier that enforces general relativity's Hamiltonian constraint generating diffeomorphisms in the radial direction, and that the nonlinear self-interaction of gravity can be modeled by higher powers of mass $a$.  In section \ref{sec:simple}, we show that the feedback of fluids, more particularly the cases of conformal fluid and causal limit fluid, on the geometry can indeed be encoded in such a Hamiltonian. 

Pushing this line of thought further, we show that the Kawai-Yokokura (KY) solution \cite{KY1,KY2,KY3,KY4,KY5} for gravity coupled to renormalized quantum matter can also be recovered by a similar effective Hamiltonian, quadratic in the mass function $a$.
The KY metric is a non-perturbative solution (in $\hbar$) to the semi-classical Einstein equations with renormalized quantum matter fields, including the effect of 4D conformal anomaly. It does not have a classical singularity at $r=0$, although it acquires a near-Planckian curvature. It does not have an event horizon but nevertheless fits with the Schwarzschild metric slightly away from the Schwarzschild radius (by a mesoscopic distance). It can be understood as a gravitational condensate consisting of many excited quanta and reproduce the Bekenstein-Hawking entropy from the bulk quantum matter fields \cite{Y1,Y2}.
%The KY metric is a non-perturbative solution to the semi-classical Einstein equations with renormalized quantum matter fields, where the feedback of the quantum matter on the geometry is consistently taken into account by a 4D conformal anomaly calculation. As a result, it does not have a singularity at $r=0$, although it acquires a curvature in $1/\hbar$. It does not have an event horizon but nevertheless fits with the Schwarzschild black hole metric slightly away from the Schwarzschild radius (by a mesoscopic distance) \cite{KY1,KY2,KY3,KY4,KY5}. 
%
%many matter fields in $\hbar$, consistent with the 4D conformal anomaly, without a horizon or a large singularity \cite{KY1,KY2,KY3,KY4,KY5}. It can be understood as a gravitational condensate consisting of many excited quanta and reproduces the Bekenstein-Hawking entropy
%from the bulk structure \cite{Y1,Y2}. 
This shows that the effective Hamiltonian approach allows to account for semi-classical and non-perturbative quantum effects of both matter fields and geometry on spherical compact self-gravitating objects without detailed theory of the microscopic origin of those quantum corrections. 

%to develop a more specific but more explicit analysis, we consider a simple and interesting effective Hamiltonian that can be applied to a certain class of configurations (Sec.\ref{sec:simple}).

%where $\rhoeff\equiv 4\pi r^2 (-\mathcal{J}^{\mathrm{eff}~t}_t)$ is the 1D energy density ($t$: Killing time), and the coefficient functions $c_i(r)$ characterizes the system of interest. This is reminiscent of a Ginzburg–Landau Hamiltonian with the mass $m(r)\equiv \frac{a(r)}{2G}$ as an order parameter (except for a non-trivial gravitational factor $b(r)$), a nonlinear interaction with respect to $G$, and so on. We study the possibilities of this Hamiltonian without its microscopic derivation. 

%By investigating several examples, we first find that this Hamiltonian can represent a type of configurations that do not have a specific length scale, such as electromagnetic field, conformal fluid, causal-limit fluid (Sec.\ref{sec:example}).
%
%A particularly interesting one is a quantum black hole model (\textit{KY metric}): a non-perturbative solution to the semi-classical Einstein equation with many matter fields in $\hbar$, consistent with the 4D conformal anomaly, without a horizon or a large singularity \cite{KY1,KY2,KY3,KY4,KY5}. It represents a gravitational condensate consisting of many excited quanta and reproduces the Bekenstein-Hawking entropy from the bulk structure \cite{Y1,Y2}. We construct an effective theory for KY metric. 
%It should hold the potential to bridge semi-classical gravity and quantum gravity including matter fields. 

Finally, we give the general conditions on the effective Hamiltonian so as to keep a smooth singularity-free metric, even at $r=0$. The exploration of such singularity-free Hamiltonians allows us to identify an improved version of the KY metric, valid for all $r$ (inside and outside the black hole), with a smooth transition from a bulk with constant negative Ricci scalar to the standard Schwarzschild solution outside.

%As a natural application, we use the Hamiltonian \eqref{Heff} and explore dynamics that resolves singularities (Sec.\ref{sec:regular}). Using the general conditions for the configuration $(a(r), b(r))$ to be non-singular at $r=0$, we construct an effective Hamiltonian that generates regular spacetime. It leads to, as the simplest one, a flat/de Sitter/anti-de Sitter spacetime, as expected. As the next simpler one, a configuration with a constant Ricci scalar but a non-trivial regular structure is obtained, and its long-range limit coincides with the KY metric.  Therefore, the regular KY metric can be a candidate for quantum black holes with the potential to bridge semi-classical gravity and quantum gravity including matter. 

%such an effective Hamiltonian describes the effective dynamics of a more complete version of KY metric, and should have some clue to microscopic dynamics connecting quantum matter and quantum gravity. 

%We finally discuss future possibilities of the effective action: generalization to time-dependent cases, effect of self-interaction in the mass $m(r)$, %microscopic derivation/understanding of the effective Hamiltonian, 
%and transition between semi-classical and quantum gravity inside the regular KY metric (Sec \ref{sec:dis}). 
%The physical meaning, microscopic derivation?? Jacobi's action? BCS-like? 

This formalism, although very simple in its conceptual framework and practical implementation, seems versatile enough to explore the dynamics of the geometry consistently coupled with renormalized quantum fields with gravitational feedback and the dynamics beyond general relativity, at least for spherically symmetric static space-times. Indeed it allows for a systematic exploration of modified Schwarzschild black holes, possibly singularity-free, together with their matter content on both sides of the Schwarzschild radius.

%%%%%%%%%%%%%%%%%%%%%%
\section{Spherically Symmetric Spacetimes}
%\section{Simple gravity action}
\label{sec:gravity}
%%%%%%%%%%%%%%%%%%%%%%

%%%%%%%%%
\subsection{Pure gravity}
\label{sec:setup}
%%%%%%%%%

Let us consider spherically-symmetric static space-times, with the following line elements:
\be
\label{eqn:THEMETRIC}
\rd s^2
=
-f(r)b(r)^2 \rd t^2 +f(r)^{-1}\rd r^2+r^2\rd \Omega^2,
\ee
with  $f(r)=1-a(r)/r$. The function $a(r)$ directly gives the Misner-Sharp mass $m(r)=a(r)/2G$ of the spatial ball of radius $r$.

Such metrics  obviously represent very simple, non-evolving space-times. They are nonetheless physically relevant as the generic non-rotating equilibrium configurations of gravitational collapse, for instance for (non-rotating) stars and black holes, as long as the anisotropic instabilities do not dominate the dynamics. It is true that the non-rotating condition is a very stringent condition, reducing the physical applicability of this class of metrics. Nevertheless studying generalizations of the Schwarzschild metric remains an enlightening exercise to explore the non-linear behavior of gravity coupled to dense classical and quantum matter.

To compute the reduced Einstein-Hilbert action $S_{\mathrm{EH}}$ for this class of metrics, we use the Gauss-Codazzi equation for the decomposition of the 4D Ricci scalar in terms of the intrinsic and extrinsic curvature tensors associated to the foliation by $t$-constant hypersurfaces, with normal vector $n_\mu \rd x^\mu =-\sqrt{-g_{tt}}\rd t$:
\begin{equation}
\label{RR}
R={}^{(3)}R+K_{\mu\nu}K^{\mu\nu}-K^2-2 \nabla_\mu (\nabla_n n^\mu-n^\mu \nabla_\nu n^\nu),
%R_{\mathrm{Ricci}}={}^{(3)}R_{\mathrm{Ricci}}+K_{\mu\nu}K^{\mu\nu}-K^2-2 \nabla_\mu (\nabla_n n^\mu-n^\mu \nabla_\nu n^\nu),
\end{equation}
where $h_{\mu\nu}\equiv n_\mu n_\nu+g_{\mu\nu}$  is the induced 3D metric and $K_{\mu\nu}\equiv h_\mu{}^\alpha h_\nu{}^\beta \nabla_\alpha n_\beta$ is the extrinsic tensor, with trace $K\equiv K^\mu{}_\mu$ \cite{Poisson}.
For any spherical static metric, we have $K_{\mu\nu}=0$ and $\nabla_\mu n^\mu=0$, so that the expression above reduces to
%Therefore, the formula \eqref{RR} reduces to 
\begin{equation}\label{RR2}
R={}^{(3)}R -2 \nabla_\mu \alpha^\mu,
%R_{\mathrm{Ricci}}={}^{(3)}R_{\mathrm{Ricci}} -2 \nabla_\mu \alpha^\mu,
\end{equation}
where $\alpha^\mu \equiv \nabla_n n^\mu$.
For the line element \eqref{metric}, the 4D volume element is $\sqrt{-g}=r^2 b \sin \theta$, the 3D Ricci scalar gives ${}^{(3)}R=\frac{2}{r^2}\dot a$, and the acceleration is:
\be
\alpha^\mu \p_\mu=\frac{\partial_r (-g_{tt})}{2b^2}\p_r
\,.
\ee
We can then compute the Einstein-Hilbert action:
\begin{align}
S_{\mathrm{EH}} &\equiv \frac{1}{16\pi G} \int_{\mathcal{M}} \rd^4 x \sqrt{-g} R
\nonumber \\
%&=\frac{1}{16\pi G} \int_{\mathcal{M}} d^4 x \left(\sqrt{-g}~{}^{(3)}R %-2\partial_\mu (\sqrt{-g}\alpha^\mu) \right)
%\nonumber \\
%&=\frac{1}{16\pi G} 4\pi \int^{t_2}_{t_1} dt \int dr \left(2 N \dot a -2 \partial_r (r^2 N\alpha^r) \right) \nonumber \\
&=
\frac{l_0}{2 G}
\int^{r_2}_{r_1} \rd r \left[b \dot a -\partial_r \left(\frac{r^2 \partial_r (-g_{tt})}{2b} \right) \right],
\label{EH}
\end{align}
where $l_0\equiv t_2-t_1$ is the time interval. Here, the surface term (up to the $l_0$ factor) can be expressed as 
\begin{equation}
\mathcal{Q}_t=\frac{\kappa(r)A(r)}{8\pi G},
\end{equation}
where $A(r)\equiv 4\pi r^2$, $\kappa(r)=\sqrt{-g_{tt}(r)}\alpha(r)$ is the surface gravity at $r$, and $\alpha(r)\equiv|\alpha_\mu \alpha^\mu|^{1/2}$ is the proper acceleration. This agrees with the on-shell Noether charge for time diffeomorphisms (or equivalently, the Komar charge for the Killing vector $\p_t$) \cite{Compere}.
Moreover, it is well-knwown that it is related to the entropy in black hole thermodynamics\footnotemark.
\footnotetext{
More precisely, by using a path-integral representation of the density of states $\nu$, the entropy can be obtained as $\mathcal{S}=\ln \nu\approx \frac{1}{\hbar} l_{BH} \mathcal{Q}_t(a_0)=\frac{A(a_0)}{4 \hbar G}$, where $a_0$ is the Schwarzschild radius, $l_{BH}=\hbar\beta_{BH}=\frac{2\pi}{\kappa(a_0)}$ is the periodic (Euclidean) time \cite{IyerWald}. It can also obtained as the adiabatic-invariant and integrable Hamiltonian of (dynamical) horizons for ``thermal time flow" generated by $\hbar\beta_{BH}\partial_t$ \cite{PY}.
}

Dropping the surface term and the factor $l_0$, we get the reduced gravity action for spherically-symmetric static space-times \eqref{Sg}: 
\begin{equation}
S_{\mathrm{g}}[a,b]
=
\frac{1}{2G}\int^{r_2}_{r_1} \rd r \,b(r)\dot a(r)
\,.
\end{equation} 
%
%Taking the variation of $(a,b)$ in \eqref{Sg}, we have 
%\begin{equation}\label{d_Sg}
%    \delta S_{\mathrm{g}} = \frac{1}{2G}\int^{r_2}_{r_1} dr ( \dot a \delta b - \dot b \delta a + \frac{d}{dr}(b \delta a)).
%\end{equation}
%For $\delta a|_{r_1,r_2}=0$, 
%Imposing $\delta S_{\mathrm{g}}=0$ with an appropriate boundary condition, the vacuum equations of motion are obtained as $\dot a =0$ and $\dot b=0$, leading to the solution $a(r)=\mathrm{const}.=a_0$ and $b(r)=\mathrm{const}.=b_0$. The metric \eqref{metric} becomes the Schwarzschild metric with the ADM mass $\frac{a_0}{2G}$:
%
Thus $a$ and $b$ are canonically conjugate variables. Moreover, the Hamiltonian vanishes so that $a$ and $b$ are constants of motion. Indeed, the equations of motion are simply $\dot a=\dot b =0$. So, classical solution for pure gravity are given by constants $a=a_0$ and $b=b_0$.
Setting $b_0$ to 1 (e.g. by rescaling the $t$ coordinate), we recognize the Schwarzschild solution:
\begin{equation}
\label{Sch}
\rd s^2
=
-\left(1-\frac{a_0}{r}\right)dt^2 + \left(1-\frac{a_0}{r}\right)^{-1}dr^2 + r^2 d\Omega^2
\,.
\end{equation}
Note that this can be written as $S_{\mathrm{g}}= \frac{1}{2G}\int \rd a\, b$, %(a form of Jacobi's action \cite{Landau_MC,Brown_York1})
which is manifestly invariant under diffeomorphism $r\to \tilde r=\tilde r(r)$.

\subsection{Effective Hamiltonian}
\label{sec:effH} 
%%%%%%

%To get a rough idea of what $\Heff$ in \eqref{Seff} is, let us start by examining some examples of classical matter fields.

%We here consider two types of $\Heff$ motivated by the previous examples.  

To go beyond pure gravity and general relativity, a natural direction of investigation is to simply explore the possibility of a non-vanishing Hamiltonian for our canonical pair $(a(r),b(r))$.
Indeed, under the two assumptions that we keep focussing on the static spherically symmetric equilibirum configurations (and thus do not look into out-of-equilibrium dynamics), and that the relevant physical degrees of freedom describing spherically symmetric spacetime remain\footnotemark{} the mass $a$ and lapse factor $b$, the only way to generate non-trivial profiles $(a(r),b(r))$ (in terms of the radial coordinate) is to introduce a non-vanishing Hamiltonian, as long as the studied effects can be modeled by a Lagrangian or Hamiltonian action principle.
\footnotetext{
Let us point out that this assumption of keeping the canonical pair $(a(r),b(r))$ excludes higher derivative terms in the Lagrangian, since those would generate extra degrees of freedom, formalized as higher momenta in the Hamiltonian formulation (see e.g. \cite{Woodard:2015zca}). In particular, introducing higher curvature terms in the Einstein-Hilbert action, such as $R^2$ or $R_{\mu\nu}R^{\mu\nu}$, would enrich our reduced gravitational phase space with extra canonical pairs.
}

There are two main sources that modify the pure-gravity dynamics of general relativity.
First, modified-gravity corrections can originate from both IR and UV effects creating new physics beyond general relativity. They would modify the Einstein-Hilbert action and the resulting Einstein equations, thus ineluctably generating an effective Hamiltonian.
The second source is matter. General relativity couples geometry to matter, and matter necessarily curves space-time, thus leading to non-trivial configurations of the mass and lapse. Technically, integrating over the matter fields, their classical and quantum degrees of freedom, will yield an effective action and Hamiltonian for the metric components $a$ and $b$.
These two sources of effective Hamiltonians are in fact intimately intertwined. Indeed, modified gravity corrections often come from extra fields added to general relativity (e.g. \cite{DeFelice:2010aj}) or can be interpreted a posteriori as the interaction of geometry with new effective fields (e.g. \cite{Riegert:1984kt}).

We thus introduce a general ansatz for the dynamics of spherically-symmetri static space-times \eqref{Seff0}:
\begin{equation}
\label{Seff}
    S_{\mathrm{eff}}[a,b]=\int^{r_2}_{r_1} dr \left(\frac{b(r) \dot a(r)}{2G}-\Heff[a,b,r]\right)\,.
\end{equation}
Here, the effective Hamiltonian $\Heff[a,b,r]$ obviously depends on the mass $a$ and lapse factor $b$, but can also depend explicitly on the radial coordinate $r$ (here the equivalent of time-dependent Hamiltonian). 

Due to the role of the field $b$ as a lapse factor, but also as the 4-volume density, it is natural to expect a power law scaling of the effective Hamiltonian in $b$, thus $\Heff[a,b,r]\propto b^\varsigma$.
Let us call such Hamiltonians of weight or type $(\varsigma)$, since the exponent $\varsigma$ should indeed reflect how it scales with the density.

Let us have a look at the simplest possibilities:
\begin{itemize}

\item {\bf Type (-1)~:}
Let us consider the case when the Hamiltonian scales as the inverse of $b$:
\begin{equation}
\label{Heff_I}
\Heff(a,b,r)=\frac{Q(a,r)}{b}\,. 
%\Heff^{(I)}(a,b,r)=\frac{Q(a,r)}{b}. 
\end{equation}
The equation of motion \eqref{Heq} becomes 
\begin{equation}\label{Heq_I}
\dot a = - 2G Qb^{-2}
%\dot a = - 2G \frac{Q(a,r)}{b^2}
\,,\quad
b\dot b = - 2G \partial_a Q
\,.
%b\dot b = - 2G \frac{\partial Q(a,r)}{\partial a},
\end{equation}
$a$ and $b$ are inextricably coupled. We can nevertheless plug the 2nd equation into the 1st in order to obtain a closed differential equation for the mass $a$:
\begin{equation}\label{eqI}
    \ddot{a}+\dot a^2 \partial_a \log Q-\dot a \p_r \log Q=0. 
\end{equation}
This is generically a highly non-linear and non-trivial second-order differential equation.

On the one hand, when $Q$ does not depend on $a$, the Hamiltonian simply consists of a potential in $b$.  For instance, if $\pp_a Q=\pp_r Q=0$, say $Q=1$, then the solution is a constant $b=b_0$ and a linear $a=\alpha_0 r +a_0$. This linear dependence of the mass function in the radius is definitely an interesting feature.
This linear slope for the mass function occurs for conformal fluids and causal-limit fluids, as we will see in section \ref{sec:example}, with slopes respectively $\alpha_0=3/7$ and $\alpha_0=1/2$, but these metrics do not have constant lapse factor $b$.
The linear $a(r)$ and constant $b(r)$ is actually realized by relativistic stars with a scale-invariant equation of state, which lead to arbitrary slope coefficient $\alpha_0$ depending on the parameters entering the equation of state \cite{Alho:2023mfc}. These make for intriguing black hole mimickers. 

On the other hand, a physically-relevant non-trivial choice of $Q$ is given by the coupling of a massless scalar field to the geometry, leading to the Janis-Newman-Winicour solution, as we will review in the next section.

\item {\bf Type (0)~:}
Let us consider the case when the Hamiltonian does not depend in $b$, that is $\pp_b \Heff=0$.
The equation of motion are then $\dot a=0$ and $\dot b=-2G\pp_a \Heff$.
This means that we have a constant Misner-Sharp mass, but a non-trivial lapse factor $b(r)$ whose precise evolution depends on the specific choice of $\Heff$, with the resulting metric reading:
\begin{align}\label{metric_type0}
\rd s^2=&
-b(r)^2\l(1-\f{a_0}r\r)\rd t^2
\nn\\
&+\l(1-\f{a_0}r\r)^{-1}\rd r^2
+r^2\rd \Omega^2
\,.
\end{align}
The energy density is zero but the pressure can be finite as one can see directly from the explicit expressions of the Einstein tensor components: 
\begin{align}
&G^t{}_t=0
\,,\quad
G^r{}_r=\frac{2(r-a_0)}{r^2}\frac{\dot b}{b},
\nonumber\\
&G^\theta{}_\theta
=
\frac{1}{rb}\left[(1+\frac{a_0}{2r})\dot b+(r-a_0)\ddot{b}  \right]
\,.
\end{align}
These violate the dominant energy condition in all directions, and this type of space-time metric does not seem to be physically grounded, at least, from a thermodynamic perspective, unless $b(r)$ is constant.
%
%\footnote{Note that, considering conformal matter fields and applying the Weyl anomaly \eqref{anomaly} to the trace part of the equation \eqref{Einstein_eff}, one can check that the metric \eqref{metric_type0} with $b(r)=e^{\sqrt{\frac{3}{16\pi l_p^2 c_W}} r}$ is a solution for $r\gg l_p$, although its physical meaning is not clear.\YY{If this footnote is needed, I would move it to some section. Otherwise, I would remove it.}}

\item {\bf Type (+1)~:}
Another simple scaling is the case when the Hamiltonian is linear in $b$. In fact, this could be considered as the most natural case, since it means that the Hamiltonian scales as a density of weight 1. We write
\begin{equation}
\label{Heff_II}
    \Heff[a,b,r]=b \rhoeff[a,r], 
\end{equation}
inducing the following equations of motion,
\begin{equation}
\label{Heq_II}
\dot a = 2G \rhoeff[a,r]
\,, \quad
\frac{\rd}{\rd r} \log b = - 2G \partial_a \rhoeff
\,.
\end{equation}
As the first equation is a decoupled differential equation solely on the mass $a(r)$, this system is much simpler to solve than the type $(-1)$. Once the evolution of the mass is integrated, one can plug it into the $b$-equation and obtain the lapse evolution. 

As we will see in the next section, it turns out that the cosmological constant, Maxwell fields and non-linear electrodynamics, lead to this type of effective Hamiltonian, making it relevant and appealing.

\end{itemize}

%In the next section, we consider various types of basic matter fields coupled to general relativity, and derive explicitly the effective Hamiltonian that they induce for the geometrical sector.

%(See the next subsection for $\rhoeff$.)

%%%%%%%%%%%%%%%%%%%%%%%%%%%%%%%%%%%%%%%%%
\section{Classical Matter Fields}
\label{sec:classical}
%%%%%%%%%%%%%%%%%%%%%%%%%%%%%%%%%%%%%%%%%

We explore the coupling of matter to the geometry. We consider various types of basic matter fields coupled to general relativity, and derive explicitly the effective Hamiltonian that they induce for the geometrical sector. The logic is simple: we compute the action and equations of motion for the coupled system restricted to spherically-symmetric static configurations, then integrate out the matter field and focus on the resulting dynamics of the geometry.

%%%%%%
\subsection{Cosmological constant}
\label{sec:cosmo}
%%%%%%

The simplest example is the cosmological constant $\Lambda$. The 4-volume term reads:
\begin{align}
S_\Lambda
=
-\frac{\Lambda}{8\pi G}\int \rd^4 x\, \sqrt{-g} 
=
-\frac{\Lambda l_0}{2G}\int \rd r\, r^2 b
\,.
\end{align}
Dropping the $l_0$ factor and adding this term to the pure gravity action \eqref{Sg} gives the full reduced Einstein-Hilbert action:
\begin{equation}
S_{\mathrm{tot}}^{(\Lambda)}
=
\int \rd r\,
\left(
\frac{b\dot a}{2G}-\frac{\Lambda r^2 b}{2G}
\right),
\end{equation}
leading to the effective Hamiltonian: 
\begin{equation}
\label{Heff_L}
\Heff^{(\Lambda)}(b,r)=\frac{\Lambda r^2 b}{2G}.
\end{equation}
The equations of motion are $\dot b=0$ and $\dot a=\Lambda r^2$.
Their solution is, of course, given by the (anti) de Sitter spacetime
%, depending on the sign of $\Lambda$
(after fixing the integration constants for $a$ and $b$ to 0 and 1, respectively),
\begin{equation}\label{metric_dS}
    ds^2=-\left(1-\frac{\Lambda r^2}{3}\right)dt^2+\left(1-\frac{\Lambda r^2}{3}\right)^{-1}dr^2+r^2 d\Omega^2.
\end{equation}
%which satisfies the regularity conditions \eqref{cond_a} and \eqref{cond_b}.

%%%%%%
\subsection{Massless scalar field}
\label{sec:scalar}
%%%%%%

We now consider a massless scalar field, still in the spherically-symmetric static case:
\begin{align}
\label{S_phi}
S_\phi
&=-\frac{1}{2}\int \rd^4 x\, \sqrt{-g}g^{\mu\nu}\partial_\mu \phi \partial_\nu \phi
\nonumber\\
&= -2\pi l_0 \int \rd r\, r^2 b f \dot \phi ^2,
%\equiv l_0 \int dr L_\phi, 
\end{align}
where the field $\phi$ depends only on $r$.
%the metric \eqref{metric} is applied. 
Putting this term together with the pure gravity term gives the following total action:
\be
S_{\mathrm{tot}}^{(\phi)}
=
\int \rd r\,
\left(
\frac{b\dot a}{2G}-2\pi r^2 b f \dot \phi^2
\right)\,.
%\nonumber\\
\ee
We perform a Legendre transformation on the scalar field, computing its canonical momentum,
\be
p_\phi
%:=\frac{\partial L_\phi}{\partial \dot \phi}
=-4\pi r^2 b f \dot \phi
\,,
\ee
and writing the action in its Hamiltonian form,
\be
S_{\mathrm{tot}}^{(\phi)}
=
\int \rd r\,
\left( \frac{b\dot a}{2G}+p_{\phi} \dot \phi+\frac{p_{\phi}^2}{8\pi r^2 b f} \right)
\,.
\ee
The equations of motion for the matter sector directly imply that the momentum is conserved, $\dot  p_{\phi}=0$. This allows  to trivially integrate out the matter sector and obtain an efective action for the geometry:
\be
S_{\mathrm{eff}}^{(\phi)}
=
\int \rd r\,
\left( \frac{b\dot a}{2G}+\frac{p_{\phi}^2}{8\pi r^2 b f} \right)
\,,
\ee
where the scalar field momentum $p_{\phi}$ now plays the role of a coupling constant in front of the effective Hamiltonian\footnotemark,
\begin{equation}
\label{Heff_phi}
\Heff^{(\phi)}[a,b,r]
=
\frac{-p_{\phi}^2}{8\pi r^2 b \left(1-\frac{a}{r}\right)}
\,.
\end{equation}
\footnotetext{
This is a local function of $r$, obtained by integrating out the scalar field, but it contains non-local information in the sense that the value of the constant $p_\phi$ is determined by a boundary condition.
}
This is an effective Hamiltonian of type (-1), with an inverse scaling in the lapse factor $b$. Following the conventions \eqref{Heff_I} of the previous section, we have $Q[a,r]={-p_{\phi}^2}/{8\pi r(r-a)}$, leading to the following equations of motion, from \eqref{eqI}:
\begin{equation}\label{eq_sc1}
    r(r-a)\ddot{a}+r \dot a^2+(2r-a)\dot a=0
    \,,
\end{equation}
\begin{equation}\label{eq_sc2}
    \frac{d}{dr} b^2 = \frac{G p_{\phi}^2}{2\pi r(r-a)^2}
    \,.
\end{equation}
The differential equation in $a$ is non-linear and highly non-trivial.
Nevertheless, as expected, one can show explicitly that these equations are solved by the well-known Janis-Newman-Winicour metric \cite{JNW},
\be
\label{metric_JNW}
\rd s^2
=
-\l(\frac{1-\frac{r_-}{\newr}}{1+\frac{r_+}\newr{}}\r)^{\frac{1}{\mu}}dt^2+\l(\frac{1+\frac{r_+}{\newr}}{1-\frac{r_-}\newr{}}\r)^{\frac{1}{\mu}}d\newr^2+r^2 d\Omega^2,
\ee
where $r$ is a function of a redefined radial coordinate $\newr$:
\be
\label{JNW1}
r(\newr)^2=(\newr+r_+)^{1+\frac{1}{\mu}}(\newr-r_-)^{1-\frac{1}{\mu}}
\,,
\ee
where the two radii are defined as $r_{\pm}\equiv \frac{r_0}{2}(\mu\pm1)$ in terms of the dimensionless exponent $\mu\ge 1$ and the length scale $r_0$. The latter gives the ADM energy $r_0/2G$.

To perform a direct check that this metric solves our equations of motion, it is enough to recast the JNW metric in our choice of gauge, which gives:
\begin{align}
\label{metric_JNW_tr}
\rd s^2
=
-\l[\frac{1-\frac{r_-}{\newr}}{1+\frac{r_+}\newr{}}\r]^{\frac{1}{\mu}}\rd t^2
+\l[1+\frac{r_+}{\newr}\r]\l[1-\frac{r_-}{\newr}\r]\rd r^2+r^2 \rd\Omega^2
\,,
\end{align}
corresponding to the following metric components:
%mass and lapse  functions:
\be
\label{a_JNW}
a=r(1-f)\,,\,\,\textrm{with}\,\,
f(r)=\f{\newr^2}{(\newr+r_+)(\newr-r_-)}
\,,
\ee
\be
\label{b_JNW}
b(r)=\dot \newr(r)=\sqrt{\l(1+\frac{r_+}{\newr}\r)^{1-\frac{1}{\mu}}\l(1-\frac{r_-}{\newr}\r)^{1+\frac{1}{\mu}}}.
\ee
%\begin{equation}\label{ab_JNW}
%a(r)=r\l[1-\frac{1}{\l(1+\frac{r_+}{R(r)}\r)\l(1-\frac{r_-}{R(r)}\r)}\r]
%\,,\quad
%b(r)=\dot R(r)
%\,,\nonumber
%\end{equation}
%with
%\begin{equation}
%    \dot R^2\equiv \l(\frac{dR}{dr}\r)^2=\l(1+\frac{r_+}{R}\r)^{1-\frac{1}{\mu}}\l(1-\frac{r_-}{R}\r)^{1+\frac{1}{\mu}}
%    \,.\nonumber
%\end{equation}
Plugging these expressions in our equations of motion shows indeed that they are solved by the JNW metric, if and only if the two limit radii $r_\pm$ are related to the scalar field momentum by the condition:
\be
r_0^2(\mu^2-1)=4r_-r_+
=\f{Gp_{\phi}^2}{\pi}.
\ee
The scalar field profile is then obtained by integrating the definition of the matter momentum $p_\phi=-4\pi r^2 bf\dot\phi$. It reads
\be
\label{JNW2}
\phi(\newr)=\frac{\phi_0}{\mu}\log \l|\frac{1-\frac{r_-}{\newr}}{1+\frac{r_+}\newr{}}\r|
\,,
\ee
with the field amplitude determined by $4\pi\phi_0=-p_\phi/r_0$.
%\YY{This agrees with the scalar-field configuration in the JNW solution.}
%The constants of integration are related by the conditions, $r_{\pm}\equiv \frac{a_0}{2}(\mu\pm1)$ and $\mu^2\equiv1+16\pi G \varphi_0^2$, so that the solutions are determined by solely two independent parameters. 

As it is well-known, the JNW metric, solving the classically coupled system of scalar matter field and geometry, exhibits horizon divergences (at $\newr=r_-$) depending on the value of the exponent $\mu$.
This singular behavior will be cured by quantum effects in the KY metric reviewed in the following sections.

%%%%%%
\subsection{Maxwell field}
\label{sec:EM}
%%%%%%

We next move on to an electromagnetic field and look into Maxwell theory. Assuming $A_\mu dx^\mu =A_t(r)dt$ for a spherically-symmetric, static Maxwell field, we compute its field strength: 
\begin{align}\label{F}
{\hat F}\equiv -\frac{1}{4} g^{\mu\alpha}g^{\nu \beta}F_{\mu\nu}F_{\alpha\beta}=\frac{\dot A^2}{2b^2},
\end{align}
where $A\equiv A_t$, $F_{\mu\nu}\equiv\partial_\mu A_\nu -\partial_\nu A_\mu$, and \eqref{metric} is used. 
Then, the Maxwell action reads 
\be
\label{S_A}
S_{\mathrm{A}}=\int \rd x^4\, \sqrt{-g}\,{\hat F}
=2\pi l_0 \int dr \frac{r^2}{b} \dot A^2
\,.
%\equiv l_0 \int dr L_{\mathrm{A}},
\ee
Following the same logic as with the scalar field, we add the pure gravity term and write the full action in its canonical form, 
\begin{align}
S_{\mathrm{tot}}
=
\int dr\left( \frac{b\dot a}{2G}+p_A \dot A- H_A \right)\,,
\end{align}
with the Maxwell field momentum and Hamiltonian:
\be
p_A=4\pi \frac{r^2}{b} \dot A
\,,\quad
H_A
%=p_A \dot A  -L_A
=
\frac{p_A^2 b}{8\pi r^2}\,.
\ee
%This leads to the Hamilton equations:
%\begin{align}
%    \dot A = \frac{\partial H_A}{\partial p_A}=\frac{p_A b}{4\pi r^2},~~
%     \dot p_A = - \frac{\partial H_A}{\partial A}=0.
%\end{align}
The equations of motion imply that the Maxwell field momentum is conserved, $\dot p_A =0$, so that we can fully integrate out the Maxwell field and obtain the effective action exactly driving the geometrical sector,
\begin{align}
\label{HeffMaxwell}
S_{\mathrm{eff}}^{(A)}
=
\int \rd r\,
&\left( \frac{b\dot a}{2G}-\Heff^{(A)} \right)
\\
&\textrm{with}\quad
\Heff^{(A)}[a,b,r]
=\frac{p_{A}^2 b}{8\pi r^2}
\,,\nonumber
\end{align}
where the Maxwell field momentum now plays the role of a coupling constant for the gravitational sector. This momentum has a clear and simple physical interpretation as the electric charge of the black hole. Indeed, we can use Stokes theorem to compute the electric charge: 
\begin{align}
\mathcal{Q}
&
%=\int_\Sigma d \Sigma_\mu j^\mu
=\int_\Sigma \rd \Sigma_\mu \nabla_\nu F^{\mu\nu}
=\frac{1}{2}\int_{\p \Sigma}\rd S_{\mu\nu}F^{\mu\nu}
\nn\\
&=
\int \rd\theta \rd\phi\, \sqrt{-g}F^{tr}
=
4\pi r^2 b \frac{p_{A}}{4\pi r^2 b}=p_{A}
\,,
\end{align}
where we use the on-shell expression of the momentum $p_A$ in terms of $\dot{A}$ in the final equalities.

So the Maxwell field induces an effective Hamiltonian of type $(+1)$, linear in the lapse factor, $\Heff^{(A)}=b\rhoeff^{(A)}[a,r]$ with  the effective energy density given by:
\begin{equation}
    \rhoeff^{(A)}(r)=\frac{\mathcal{Q}^2}{8\pi r^2}.
\end{equation}
As expected, the resulting equations of motion lead back to the well-known Reissner-Nordstr\"{o}m metric. 
Indeed, since the effective Hamiltonian does not depend on $a$, the first equation of motion is $\dot{b}=0$, so that the lapse factor $b$ is constant, say $b=1$.
Next, we solve the equation of motion for the mass $a(r)$:
\begin{align}
\dot a =2G \rhoeff= \frac{G\mathcal{Q}^2}{4\pi r^2}
\quad\Rightarrow\quad
a(r)=a_0 -\frac{G\mathcal{Q}^2}{4\pi r}.
\end{align}
We identify the integration constant $a_0$ as the ADM energy $M= {a_0}/{2G}$. Hence we recover the Reissner-Nordstr\"{o}m metric:
\begin{align}
\rd s^2 &=
-f(r) \rd t^2+f(r)^{-1}\rd r^2+r^2 \rd\Omega^2
\\
&\textrm{with}\quad
f(r)
=
1-\frac{2GM}{r}+\frac{G\mathcal{Q}^2}{4\pi r^2}
\,.\nonumber
\end{align}
It turns out that we can generalize this type $(+1)$ behavior to all non-linear extensions of the electromagnetic field, as explained below.

%%%%%%
\subsection{Non-linear Electromagnetic Theory}
\label{sec:NL}
%%%%%%

Let us extend our previous analysis of the Maxwell field to non-linear extensions  and consider the general action for the electromagnetic sector \cite{Ayon-Beato:1998hmi, Dymnikova:2004zc}:
\begin{align}
\label{S_EM}
S_{\mathrm{NL}}&= \int \rd x^4\, \sqrt{-g}\,\cL(\hat F)
%\nonumber\\&
=4\pi l_0 \int \rd r\, r^2 b\cL(\hat F)
%\equiv l_0 \int dr L_{\mathrm{NL}},
\end{align}
where $\cL(\hat F)$ is an arbitrary function of the field strength $\hat F$ given earlier in \eqref{F}. 

We add the gravitational action and get the full action for the coupled system geometry plus matter:
\begin{align}
S_{\mathrm{tot}}
=
\int \rd r\,\left( \frac{b\dot a}{2G}+p_A \dot A- H_{\mathrm{NL}} \right)\,,
\end{align}
with the modified Maxwell-field momentum and Hamiltonian:
\be
p_A=\frac{4\pi r^2}{b}\dot A \pp_{\hat F} \cL
\,,\quad
H_{\mathrm{NL}}
%=p_A \dot A  -L_A
=
b\l[
\frac{p_A^2}{4\pi r^2 \pp_{\hat F} \cL}
-4\pi r^2 \cL
\r]
\,.
\nn
\ee
%This leads to the Hamilton equations:
%\begin{align}
%    \dot A = \frac{\partial H_A}{\partial p_A}=\frac{p_A b}{4\pi r^2},~~
%     \dot p_A = - \frac{\partial H_A}{\partial A}=0.
%\end{align}
Because the field strength $\hat F$ depends only on the derivative $\dot{A}$ and not directly on $A$, this Hamiltonian does depend only on the momentum $p_A$ and not on  $A$. The resulting equations of motion imply that the Maxwell field momentum is conserved, $\dot p_A =0$. Thus, as before, we can fully integrate out the electromagnetic field and obtain the effective action exactly driving the geometrical sector,
\begin{align}
S_{\mathrm{eff}}^{(NL)}
&=
\int \rd r\,
\left( \frac{b\dot a}{2G}-\Heff^{(NL)} \right)
\\
&\textrm{with}\quad
\Heff^{(NL)}[a,b,r]
=
b\l[
\frac{p_A^2}{4\pi r^2 \pp_{\hat F} \cL}
-4\pi r^2 \cL
\r]
\,,\nonumber
\end{align}
where the field momentum $p_A$ plays the role of a coupling constant for the geometry although the $p_A$-dependence is intricate generically.
This is always an effective Hamiltonian of type $(+1)$ as for the standard Maxwell field.
Moreover, since the field strength $\hat F$ does not depend on $a$, the effective Hamiltonian never also depends directly on the metric component $a$, so the lapse factor is always constant on classical solutions, $\dot b=0$, and only the dynamics of the mass $a$ is affected by the non-linear extension of the EM field.

%%%%%%%%%

%To find the effective Hamiltonian, we first define the momentum for $A$ as 
%\begin{equation}\label{p_NL}
%    p_{A}:=\frac{\p L_{\mathrm{NL}}}{\p \dot A}=4\pi r^2 b \frac{\p \cL}{\p F} \frac{\p F}{\p \dot A}=\frac{4\pi r^2}{b} \frac{\p \cL}{\p F} \dot A,
%\end{equation}
%where \eqref{F} is applied. Combining this and \eqref{F}, we can find a relation
%\begin{equation}\label{F_relation}
%    F \left(\frac{\p \cL}{\p F}\right)^2 = \frac{p_A^2}{32\pi ^2 r^4}.
%\end{equation}
%Because the left hand side is a function of $F$, this relation can determine $F$ in terms of $p_A$ and $r$. 
%We then calculate 
%\begin{align}
%    H_{NL}:=p_A \dot A  -L_{NL}
%    =b\left(\frac{p_A^2}{4\pi r^2}\frac{1}{\frac{\p \cL}{\p F}}-4\pi r^2 \cL \right).
%\end{align}
%Here, $A$ does not appear, and the equation of motion, $\dot p_A = - \frac{\p H_A^{(NL)}}{\p A}=0$, leads to $p_A=\mathrm{const.}\equiv p_{A0}$. Thus, we obtain the effective Hamiltonian:
%\begin{align}\label{Heff_NL}
%    \Heff^{(NL)}(b,r)=b\left. \left(\frac{p_A^2}{4\pi r^2}\frac{1}{\frac{\p \cL}{\p F}}-4\pi r^2 \cL \right) \right|_{p_A=p_{A0}},
%\end{align}
%which is $a$-independent. For $\cL=F$, it reduces to \eqref{HeffA}. 

Now that we have found that the simplest matter fields explicitly provide examples of effective Hamiltonian for the geometrical sector, particularly of types $(-1)$ and $(+1)$, we can turn to the general constraints of the effective Hamiltonian set by its compatibility with the Einstein equations and, in particular, with fluid matter.

%%%%%%%%%%%%%%%%%%%%%%%%%%%%%%%%%%%%%%%%%
\section{General structure of $\Heff$}
%\section{Generic Properties of Effective Hamiltonian}
\label{sec:general}
%%%%%%%%%%%%%%%%%%%%%%%%%%%%%%%%%%%%%%%%%

The previous section illustrated the logic of deriving effective Hamiltonians that drive the (radial) evolution of the geometrical sector once the dynamics of other fields (e.g. matter) has been integrated. We provided explicit solvable cases such as the coupling of gravity to a scalar field or to the electromagnetic field.
Let us investigate next the general structure of effective Hamiltonians $\Heff[a,b,r]$ and possible constraints that they must satisfy if they reflect the coupling of gravity to matter sources.

%%%%%%
%\subsection{General structures of $\Heff$}
%\label{sec:gene_H}
%%%%%%

%%%%%%
\subsection{Compatibility with Einstein equations}
\label{sec:gene_H}
%%%%%%%

%We now investigate the structure of the general effective Hamiltonian $\Heff(a,b,r)$.
%Generically, it can be complicated (see also footnote \ref{foot:H}), but the effective dynamics equation \eqref{eom} can constrain the structure to some extent. 

We are dealing with general relativity, so the dynamics generated by the effective Hamiltonian for spherically-symmetry static metrics should reflect the Einstein equations \eqref{eqn:firsteqn}. Let us investigate if this consistency check leads to constraints on possible effective Hamiltonians.

The Einstein tensor of the considered metric \eqref{metric} reads:
\begin{align}\label{G_ab}
&G^t{}_t=-\frac{\dot a}{r^2}
\,,\quad
G^r{}_r=-\frac{\dot a}{r^2}+\frac{2(r-a)}{r^2}\frac{\dot b}{b},
\nonumber\\
&G^\theta{}_\theta
=
\frac{1}{2rb}\left[(2+\frac{a}{r}-3\dot a)\dot b-b\ddot{a}+2(r-a)\ddot{b}  \right]
\,.
\end{align}
The Einstein equations $G_{\mu\nu}=8\pi G \cT_{\mu\nu}$ coupling spherically symmetric static matter to gravity then give equations of motion for $a$ and $b$,
%Using the first two equations and the effective equation of motion \eqref{eom}, we have 
\begin{align}
\dot a = 2G \rho
\,,\quad
\dot b =2G \frac{\rho+{p_r}}{2(r-a)}b,
%\dot a = 2G \rhoeff,~~\dot b =2G \frac{\rhoeff+{p_\mathrm{eff}}}{2(r-a)}b,
\end{align}
where $\rho\equiv 4\pi r^2(- \cT{}^t{}_t)$ and  $p_r\equiv 4\pi r^2 \cT{}^r{}_r$ are respectively the energy density and radial pressure averaged over the sphere of radius $r$.
Comparing to the equations of motion \eqref{Heq} for the effective Hamiltonian implies that 
\begin{align}
\label{Heff_relation}
\frac{\p \Heff}{\p b}=\rho
\,,\quad
\frac{\p \Heff}{\p a}=-\frac{\rho+p_r}{2(r-a)}b
\,.
\end{align}
This provides a direct physical interpretation of the (derivatives of the) effective Hamiltonian as generating energy density and pressure sources for the geometrical sector.
This is a consistency relation that must hold in general.
%between the effective dynamics \eqref{eom} and the effective Hamiltonian $\Heff$ in \eqref{Seff}.
In fact, all the examples of classical matter fields presented in section \ref{sec:classical} satisfy this relation.

Furthermore, for type $(+1)$ effective Hamiltonian, $\Heff=b\rhoeff[a,r]$, we clearly see the matching between the factor $\rhoeff[a,r]$ and the energy density $\rho$, thus justifying our chosen notation.
Following this insight, given any effective Hamiltonian $\Heff$, we will write $\rhoeff$ and $\peff$ for the functions resulting from the derivatives of the Hamiltonian according to \eqref{Heff_relation}, and will refer to them as the effective energy density and effective pressure.

Still focussing on effective Hamiltonian of type $(+1)$, thus $\Heff=b\rhoeff[a,r]$,  the compatibility with the Einstein equations implies that
\be
b\frac{\p \rhoeff}{\p a}=\frac{\p \Heff}{\p a}=
-\frac{\rhoeff+\peff}{2(r-a)}b\,,
\nn
\ee
or written in a flatter fashion:
\begin{equation}\label{eq_of_state}
    \peff(a,r)=-\rhoeff(a,r)-2(r-a)\frac{\p \rhoeff(a,r)}{\p a}.
\end{equation}
This relation between radial pressure and energy density is an equation of state that holds generally for effective Hamiltonians of type $(+1)$. In particular, for thermodynamical configurations, e.g. highly excited configurations, which 
naturally have both positive energy density and pressure \cite{Landau:1980mil,Goldstein:2005aib}, this gives a condition of the sign of the derivative of the effective energy density $\rhoeff$ with respect to the mass $a$:
\begin{equation}
\label{rhoeff_rel}
\rhoeff+\peff>0 \Leftrightarrow \frac{\p \rhoeff(a,r)}{\p a}<0
\,,
\end{equation}
where we have used the condition that the spacetime is static: $r>a(r)$.
\subsection{Linear Equations of State}
%\subsubsection*{Equation of state}
%\label{sec:gene_H}
%%%%%%

Let us study in more detail  the widely used case of a fluid with linear barotropic equations of state,
\begin{equation}\label{eta_eq}
    \peff=\frac{2-\eta}{\eta} \rhoeff,
\end{equation}
where $\eta$ is a constant parameter characterizing the fluid.
This can be applied to both a locally isotropic fluid ($\cT{}^r{}_r=\cT{}^\theta{}_\theta$ \cite{Nilsson:2000zf}) and an anisotropic material ($\cT{}^r{}_r\neq \cT{}^\theta{}_\theta$).
Requiring the dominant energy condition $\peff\leq \rhoeff$, and the positivity of $\peff$ and $\rhoeff$, imply that $\eta$ must satisfy 
\begin{equation}\label{eta}
    1\leq \eta < 2\,.
\end{equation}
Inserting this equation of state in the condition \eqref{Heff_relation} on the derivatives of the effective Hamiltonian yields:
%, we have $\frac{\p \Heff}{\p a}=-\frac{\rhoeff}{\eta(r-a)}b$. We connect this to the first one and obtain an non-trivial relation: 
\begin{equation}
\eta (r-a) \frac{\p \Heff}{\p a}=-b \frac{\p \Heff}{\p b}
\,.
\end{equation} 
This means that $\Heff$ can depend on $a$ and $b$ only through a single variable $c$ given by
\begin{equation}
\label{Heff_c}
\Heff[a,b,r]=\cHeff[c,r]
\,\,\textrm{with}\quad
c\equiv b\left(1-\frac{a}{r}\right)^{\frac{1}{\eta}}\,,
\end{equation}
without any specific constraint on the way it depends on the radial coordinate $r$.
This allows to recover in a straightforward fashion the various examples of matter explored in the previous section.
Indeed, for the cosmological constant or a Maxwell field, one has $\cT^t{}_t=\cT^r{}_r$, corresponding to $\eta=\infty$, implying that the effective Hamiltonian should not depend on the field $a$ but solely on $b$. This is indeed consistent with what was derived earlier in \eqref{Heff_L} and \eqref{HeffMaxwell}. As for the massless scalar field, one has $-\cT^t{}_t=\cT^r{}_r$, indicating $\eta=1$ and thus implying that the effective Hamiltonian should be a function of $c=b(1-a/r)$. This is perfectly consistent with our earlier derivation of the effective Hamiltonian \eqref{Heff_phi} induced by a massless scalar field, which we recall here:
\be
\Heff^{(\phi)}
=
\frac{-p_{\phi}^2}{8\pi r^2 b \left(1-\frac{a}{r}\right)}
\,.
\nn
\ee
%
%This should be useful in considering the effective Hamiltonian for a system with the equation of state \eqref{eq_of_state}. Note that for massless scalar field \eqref{S_phi}, we have $-T^t{}_t=T^r{}_r$, indicating $\eta=1$ and $\Heff^{(\phi)}(c,r)=\frac{-p_{\phi 0}^2}{8\pi r^2 c}$ with $c=b(1-\frac{a}{r})$ from \eqref{Heff_phi}; and for Maxwell theory \eqref{S_A}, we have $T^t{}_t=T^r{}_r$, meaning $\eta=\infty$ in \eqref{eta_eq}, and \eqref{HeffA} shows $\Heff^{(A)}(c,r)=\frac{p_{A0}^2 c}{8\pi r^2}$ with $c=b$. 
%

Now that we have the basic definition, examples and consistency equations for effective Hamiltonians for the dynamics of spherically-symmetric static metrics, we shall investigate the physics generated by such Hamiltonians.

%that can represent either locally isotropic fluid ($\Jeff{}^r{}_r=\Jeff{}^\theta{}_\theta$) or anisotropic material  ($\Jeff{}^r{}_r\neq\Jeff{}^\theta{}_\theta$).

%%%%%%%%%%%%%%%%%%%%%%%%%%%%%%%%%%%%%%%%%%%%%%%%%%
\section{Basic Hamiltonian Ansatz}
\label{sec:simple}
%%%%%%%%%%%%%%%%%%%%%%%%%%%%%%%%%%%%%%%%%%%%%%%%%%
%\YY{We have reviewed so far the definition, examples, and consistency equations for the effective Hamiltonian of spherically symmetric static spacetime. From now, we will assume such a Hamiltonian and examine the physics.} 

%%%%%%%%%%%%%%%
\subsection{Polynomial ansatz in powers of mass}
\label{sec:Heffansatz}
%%%%%%%%%%%%%%%

Let us now investigate in a more systematic way the dynamics generated by effective Hamiltonians. In particular, we will focus on type $(+1)$ Hamiltonian, i.e. that depends on linearly on the lapse factor $b$. This is strongly motivated by the roles of $b$ as the 4-volume density and as the Lagrange multiplier for radial diffeomorphisms.  It is then natural to look at an expansion of the energy density in powers of $a$:
%
%We now propose a simple effective Hamiltonian. As seen above, the type-II Hamiltonian \eqref{Heff_II} is simpler. Then, a natural and simple ansatz of $\rhoeff(a,r)$ is a polynomial in $a$. Therefore, the effective Hamiltonian we will focus on is the form of \eqref{Heff} with \eqref{rhoeff}:
\begin{align}
\label{Heff_P}
\Heff[a,b,r]&=b \rhoeff[a,r]\,, \quad\textrm{with}\quad
\\
%\label{rhoeff}
2G\rhoeff[a,r]&=c_0(r)+c_1(r)a(r)+\frac{1}{2}c_2(r)a(r)^2+\cdots.
\nonumber
\end{align}
Remembering that the physical interpretation of $\frac{a(r)}{2G}$ as the Misner-Sharp mass contained in the ball of radius $r$, this is simply an expansion of the energy density in powers of the mass. Higher powers in the mass reflect the contribution from non-linear self-interactions of the gravitational field to the energy density.

The coefficient functions $c_n(r)$ characterize the effective model and encode the non-linear dependence of the effective dynamics in the mass. In principle, they could be determined from microscopic dynamics through the exact integration over those microscopic degrees of freedom or a coarse-graining flow.

The induced equations of motion \eqref{Heq} are given by: 
\begin{align}\label{eq_a}
&\frac{d}{dr} a = c_0+c_1a+\frac{1}{2}c_2a^2+\cdots\\
\label{eq_b}
&\frac{d}{dr} \log b =-(c_1+c_2 a + \cdots). 
\end{align}
Let us keep in mind that the coefficients $c_n$ are actually functions of $r$.
The natural strategy is to try to first integrate the first non-linear differential equation \eqref{eq_a} for $a$, and then plug the identified solutions for $a$ in the equation \eqref{eq_b} for $b$.
To solve these approximatively, at least in the region with large radius $r$ compared to the Planck length $l_P\equiv\sqrt{\hbar G}$, one can expand both $a(r),b(r)$ and the coefficients $c_n(r)$ in power series in $r^{-k}$ and determine the expansion coefficients recursively.

\medskip

Let us start by considering the case when $\rhoeff[a,r]$ does not depend at all on $a$,
\be
\left|
\begin{array}{l}
\rhoeff[a,r]=c_0(r)\,,
\vspace*{1mm}\\
c_n(r)=0\,\,{\rm{for}}\,\,n\geq 1\,,
\end{array}
\right.
\quad\Rightarrow\quad
\begin{array}{l}
\dot a = c_0\,,
\vspace*{1mm}\\
\dot b = 0\,,
\end{array}
%\right.
\ee
so that $b$ is constant, say $b=1$, and $a$ is simply the radial integral of $\rhoeff$. This case
%\footnotemark{}
covers a wide range of standard black holes metrics, such the Schwarzschild metric, Reissner-Nordstr\"{o}m metric, and various examples of regular black-hole metrics 
\cite{1968qtr..conf...87B,Dymnikova:1992ux, Ayon-Beato:1998hmi, Dymnikova:2004zc, Hayward:2005gi}, and so on.

\smallskip

Now going one step further in the power series, truncating $\rhoeff$ to the linear level in $a$, thus with $c_n=0$ for $n\ge2$, yields easily integrable differential equations:
\be
\dot a=c_0+c_1a
\,\Rightarrow\,\,
a
=
e^{\int^r c_1}
\bigg{[}
\int^r \rd \tilde{r}\,c_0(\tilde{r})e^{-\int^{\tilde{r}} c_1}+a_0
\bigg{]}\,,
\nn
\ee
\be
\dot b=-c_1 b
\quad\Rightarrow
b \propto e^{-\int^r \rd r\, c_1}
\,,
\ee
allowing to manufacture non-trivial lapse factors $b(r)$. 

\smallskip

Truncating the power series of $\rhoeff[a,r]$ to quadratic order in $a$, with only coefficients $c_0,c_1,c_2$, gives a Riccati differential equation, which can be dealt with using standard mathematical methods.
This truncation already defines a vast choice of effective Hamiltonians leading to various physics. As we will show below, it is rich enough, for instance, to allow for various fluids and even for the KY solution \cite{KY1} that stabilizes the black hole interior due to the non-linear coupling of renormalized quantum matter with the geometry. 

Let us conclude this preliminary analysis with the important idea of degeneracy, i.e. different Hamiltonians can lead to the same solutions for $a(r),b(r)$.
For example, let us come back to the case of constant $b$, generated by an effective Hamiltonian with no dependence on $a$, $\Heff=bc_0(r)$. It turns out that the same metric can be realized by a quadratic Hamiltonian,
\be
0=\dot b= -b(c_1+c_2 a)
\quad\Rightarrow\,\,
a=-c_1/c_2\,,
\ee
but the equation of motion for $a$ \eqref{eq_a} imposes a necessary condition on the coefficients $c_n(r)$:
\be
c_1\dot c_2-\dot c_1c_2
=
c_0c_2^2-\f12c_2c_1^2
\,,
\ee
which determines $c_0$ in terms of $c_1$ and $c_2$.
This clearly excludes the possibility that $c_1=c_2=0$ and realizes another Hamiltonian, with different coefficients $c_n(r)$, generating a solution with an arbitrary $a(r)$ and a constant $b(r)$. 
This degeneracy property means that one would need other criteria to select a particular effective Hamiltonian, either its naturalness in terms of the metric components or curvature tensors, or its direct derivation from a more fundamental action with matter coupling or modified gravity (as we have done earlier in simple cases of classical matter fields in section \ref{sec:classical}).

%This could be intuitively interpreted as an effective theory that determines the energy distribution $m(r)\equiv \frac{a(r)}{2G}$ through the non-linear self-interaction via the coupling $G$. %the 1D energy density averaged over a sphere of radius $r$ through nonlinear interactions between energies $m(r)\equiv a(r)/2G$ within $r$ via $G$. 
%
%Alternatively, this could be viewed as a Landau Hamiltonian with $a(r)$ as an order parameter (though the coefficients may depend on $r$). However, at this stage, it remains unclear how this is derived at the most fundamental level and what its physical significance is. In the following, we will assume this effective Hamiltonian and explore its possibilities. 

%%%%%%%%%%%%%%%
\subsection{Effective Hamiltonians for Fluid Matter}
%\subsection{Example of configurations for the Hamiltonian}
\label{sec:example}
%%%%%%%%%%%%%%%

%In order to investigate the properties and scope of application of the effective Hamiltonian \eqref{Heff} with \eqref{rhoeff}, we demonstrate how to use it for several examples where direct construction of the effective action is difficult.
%
%In order to investigate the property of the effective Hamiltonian $\Heff$ \eqref{Heff} with \eqref{rhoeff} and the scope of application, we demonstrate some examples, whose effective action is difficult to construct directly.

%Let us combine this effective Hamiltonian ansatz with the structure resulting from equations of state to illustrate the scope of the presented framework in cases for which direct construction of the effective action from a fundamental matter Lagrangian is not possible.
To illustrate the scope of the presented framework, let us construct the effective Hamiltonian for self-gravitating isotropic fluids, whose effective action is difficult to construct directly from a fundamental matter Lagrangian.\footnote{In Ref.\cite{Brown:1992kc}, an effective action of perfect fluid is proposed. Adding boundary terms properly and reducing it to spherically-symmetric static cases, the effective Hamiltonian becomes $b \rho$, although $\rho$ is not a function of $a$ and $r$ generically.}

%%%%%%%%%%
\subsubsection{Conformal Fluid}\label{sec:conf}
%%%%%%%%%%

A conformal fluid is characterized by the isotropic condition $T^r{}_r=T^\theta{}_\theta$ and the traceless condition $T^\mu{}_\mu=0$. In the spherically symmetric static case, these two conditions fully determine the metric \cite{Weinberg:1972kfs}: 
\begin{align}\label{metric_conf}
    ds^2=-\frac{4r}{7\err}dt^2+\frac{7}{4}dr^2+r^2 d\Omega^2
\end{align}
for $l_p\ll r\leq \err$, where $\err$ is the size of the fluid star (see Appendix A in \cite{Y2} for a heuristic derivation). This interior metric connects to the exterior Schwarzschild metric \eqref{Sch} with mass $a_0=a(\ell)$ at $r=\err=\frac{7}{3}a(\err)$.
%$=\frac{7}{3}a_0$. 
%
Comparing the metric \eqref{metric_conf} with our spherically-symmetric ansatz \eqref{metric} gives the following metric components:
\begin{equation}
\label{ab_conf}
a(r)=\frac{3}{7}r
\,,\quad
b(r)=\sqrt{\frac{r}{\err}}
\,.
\end{equation}

Let us look for an effective Hamiltonian of type $(+1)$, thus an effective energy density $\rhoeff[a,r]$, that fits with this evolution $(a(r),b(r))$. Let us assume, for simplicity, a quadratic truncation with
\begin{equation}
\label{assumption_c}
c_n(r)=0~~{\mathrm{for}}~~n\geq3
\,.
\end{equation}
Plugging the profiles $a(r),b(r)$ into the equation of motion \eqref{eq_b} for the $b$-component gives a relation between $c_1$ and $c_2$:
\be
c_1(r)=-\frac{1}{2r}-\frac{3}{7}c_2(r)r
\,.
\ee
Doing the same with the equation of motion \eqref{eq_a} for the $a$-component, in turn, gives
\be
c_0(r)=\frac{9}{14}+\frac{9}{98}c_2(r)r^2\,.
\ee
Putting these two conditions together yields:
\begin{equation}
\label{rho_conf}
2G\rhoeff
%[a,r]
=
\frac{9}{14}+\frac{9}{98}c_2r^2-\left(\frac{1}{2r}+\frac{3}{7}c_2r\right)a+\frac{1}{2}c_2a^2
\,,
\end{equation}
where $c_2(r)$ is an arbitrary function. Let's not forget (from \eqref{rhoeff_rel}) that we still need to impose $\frac{\p \rhoeff(a,r)}{\p a}<0$ to respect the positive energy condition.
%from \eqref{rhoeff_rel}. 
%and cannot be fixed by this argument. 
%
This constraint is automatically satisfied for the simplest choice $c_2(r)=0$, which reads:\footnote{Instead of setting $c_2(r)=0$, noting $\eta =3/2$ for conformal fluids, we can ask the consistency between the ansatz \eqref{Heff_P} and the condition \eqref{Heff_c} up to the linear level in $a$ and $b$ and find the consistent form of $c_2(r)$ for a large $r$.}  
\begin{equation}
\label{rho_conf0}
2G\rhoeff[a,r]
=
\frac{9}{14}- \frac{1}{2r}a
\,.
\end{equation}

As a consistent check, let us start with this effective Hamiltonian and solve the equations of motion \eqref{eq_a} and \eqref{eq_b}. We get exact solutions:
\begin{equation}
a(r)=\frac{3}{7}r+Cr^{-1/2}\,,\quad
%\approx \frac{3}{7}r
b(r)=b_0 r^{1/2}\,,
\end{equation}
where $C,b_0$ are integration constants.
This indeed gives back the desired metric \eqref{ab_conf}, either by taking $C=0$, or, in general, for $r\gg l_p$ in which the $r^{-1/2}$ term is clearly subdominant.
%
%Considering the connection to the exterior Schwarzschild metric and choosing the integration constant $b_0$ properly, we can reproduce \eqref{ab_conf} indeed. 
%
Let us underline that the space of solutions to the equations of motion induced by our effective Hamiltonian is larger than only the target metric \eqref{ab_conf}, since we have an extra term $Cr^{-1/2}$ with arbitrary parameter $C$. A natural question would be to endow it with a concrete physical interpretation in terms of fluid or gravitational dynamics. We haven't found such a physical meaning for this new term, but a possible line of investigation could be to use the analysis of the $1/r$ expansion of asymptotically flat space-time, e.g. \cite{Geiller:2022vto}.

Finally, we can apply our equation of state formula \eqref{eq_of_state} to compute the effective pressure from the effective energy density. Using $a(r)=\frac{3}{7}r$ gives
%Applying \eqref{rho_conf0} to \eqref{eq_of_state} and using $a(r)=\frac{3}{7}r$, we find
$\peff=\frac{1}{14G}=\frac{1}{3} \rhoeff$, which is consistent with $T^\mu{}_\mu=0$.

We can thus conclude that the effective Hamiltonian 
\be
\Heff^{conf}[a,b,r]=
\f{b}{4G}
\Bigg{(}
\frac{9}{7}- \frac{a}{r}
\Bigg{)}
\ee
encodes the full dynamics of the feedback of a conformal fluid on the dynamics of the geometry.

%%%%%%%%%%%%%%%%%
\subsubsection{Causal-limit Fluid}
\label{sec:causal}
%%%%%%%%%%%%%%%%%

Next, we consider a causal-limit fluid. It is defined by the isotropy condition $T^r{}_r=T^\theta{}_\theta\equiv p_{3d}$ and the equation of state $p_{3d}=\rho_{3d}(\equiv -T^t{}_t)$, which saturates the dominant energy condition and corresponds to the limiting situation in which causal propagation is established \cite{Zeldovich:1961sbr}.
As shown in e.g. Appendix A of \cite{Y2}, these two conditions determine the metric of the spherically-symmetric static configuration with size $\err$: 
\begin{equation}\label{metric_Zel}
    ds^2=-\frac{r^2}{2\err^2}dt^2+2dr^2+r^2d\Omega^2
\end{equation}
for $l_p \ll r \leq \err$, which corresponds to:
\begin{equation}\label{ab_Zel}
    a(r)=\frac{1}{2}r,~~b(r)=\frac{r}{\err},
\end{equation}
and connects to the Schwarzschild metric \eqref{Sch} with mass $a_0=a(\ell)$, at $r=\err=2a(\err)$.
%$=2a_0$.

Following the same procedure as above for the conformal fluid example, we derive compatible effective energy densities:
%for \eqref{ab_Zel} 
\begin{equation}\label{rho_Zel}
2G\rhoeff[a,r]=1+\frac{c_2}{8}r^2-\left(\frac{1}{r}+\frac{c_2}{2}r\right)a+\frac{1}{2}c_2a^2\,,
\end{equation}
where $c_2(r)$ is still an arbitrary function.
Picking the simplest choice asuming $c_2(r)=0$, we have 
\begin{equation}\label{rho_Zel0}
    2G \rhoeff[a,r]=1-\frac{a}{r}.
\end{equation}
Reversely, we can solve explicitly the equations of motion \eqref{eq_a} and \eqref{eq_b} generated by the corresponding effective Hamiltonian $\Heff=b\rhoeff$, obtaining:
\begin{equation}
a(r)=\frac{r}{2}+\frac{C}{r}\underset{r\gg l_p}\approx \frac{r}{2}
\,,\quad
b(r)=b_0 r,
\end{equation}
with two free integration constants $C$ and $b_0$, reproducing the metric components \eqref{ab_Zel} as expected for $r\gg l_p$.

Thus, the effective Hamiltonian for the geometry coupled to a causal-limit fluid is given by 
\begin{equation}
\label{Heff_causal}
\Heff^{causal}[a,b,r]
=\frac{1}{2G}b\l(1-\frac{a}{r}\r)=\frac{1}{2G}bf,    
\end{equation}
where $f(r)=1-{a(r)}/{r}$ is the usual metric factor entering the spherically-symmetric ansatz \eqref{metric}.
We do not have a simple argument for obtaining this surprisingly simple effective Hamiltonian.
Nevertheless, we point out that $bf$ is simply $\sqrt{{-g_{tt}}/{g_{rr}}}$, that is, the speed of light for radial light-like geodesics. 
%
%To speculate the meaning of this simple form, we note $b=\sqrt{-g_{2D}}=\sqrt{-g_{tt}g_{rr}}$ and $f=1/g_{rr}$ and express \eqref{Heff_causal} as $\Heff^{causal}=\frac{1}{2G}\sqrt{\frac{-g_{tt}}{g_{rr}}}$, which could correspond to a $r$-dependent speed of light in the radial direction, obtained by $ds^2=g_{tt}dt^2+g_{rr}dr^2=0$.
%
This might  be more that a mere coincidence for the spherically-symmetric causal-limit fluid, which has no characteristic properties except for the limiting behavior $\rho_{3d}=p_{3d}$, but we postpone the investigation of a deeper origin of this effective Hamiltonian to future work.\footnote{For causal-limit fluids, we have $\eta=1$ and $c=bf$ in \eqref{Heff_c}. Therefore, the simple form \eqref{Heff_causal} is automatically consistent with the condition \eqref{Heff_c}. This could be a possible origin.}

%%%%%%%%%%%%%%%%%%%%%%%%%%%%%%%%%%%%%%%%%%%%%%%%%%
\subsection{Generating Kawai-Yokokura metrics}
%\subsection{Generating the Kawai-Yokokura black hole metric}
%\subsection{Generating the KY black hole metric}
%\subsection{A Quantum Black-Hole Model: KY metric}
\label{sec:KY}
%%%%%%%%%%%%%%%%%%%%%%%%%%%%%%%%%%%%%%%%%%%%%%%%%%

%Now, let us discuss a particularly interesting example described by the effective action with \eqref{Heff} and \eqref{rhoeff}, a quantum black hole model (KY metric). Here, we will glimpse the possibility of an effective description of quantum dynamics that simultaneously represents matter and gravity, as mentioned in Sec.\ref{sec:intro}.

Now, we move on to the main application of the effective Hamiltonian framework which we wish to present in this paper. We are interested in the quantum black hole model defined by the KY metrics \cite{KY1}, which solves non-perturbatively the semi-classical Einstein equations coupling gravity to renormalized quantum matter fields. We aim to show that it can be derived from simple effective Hamiltonians, which offer new insight on the energy density profile inside the black hole and open new possibilities to explore corrections, extensions and improvements of the model.

%%%%%%%%%%%%%%%%%%%%
\subsubsection{The KY solution for Quantum Black Holes}
%\subsubsection{The KY black hole metric}
%\subsubsection{Review of KY metric}
\label{sec:review_KY}
%%%%%%%%%%%%%%%%%%%%

\begin{figure}[h]
\begin{center}
\includegraphics*[scale=0.55]{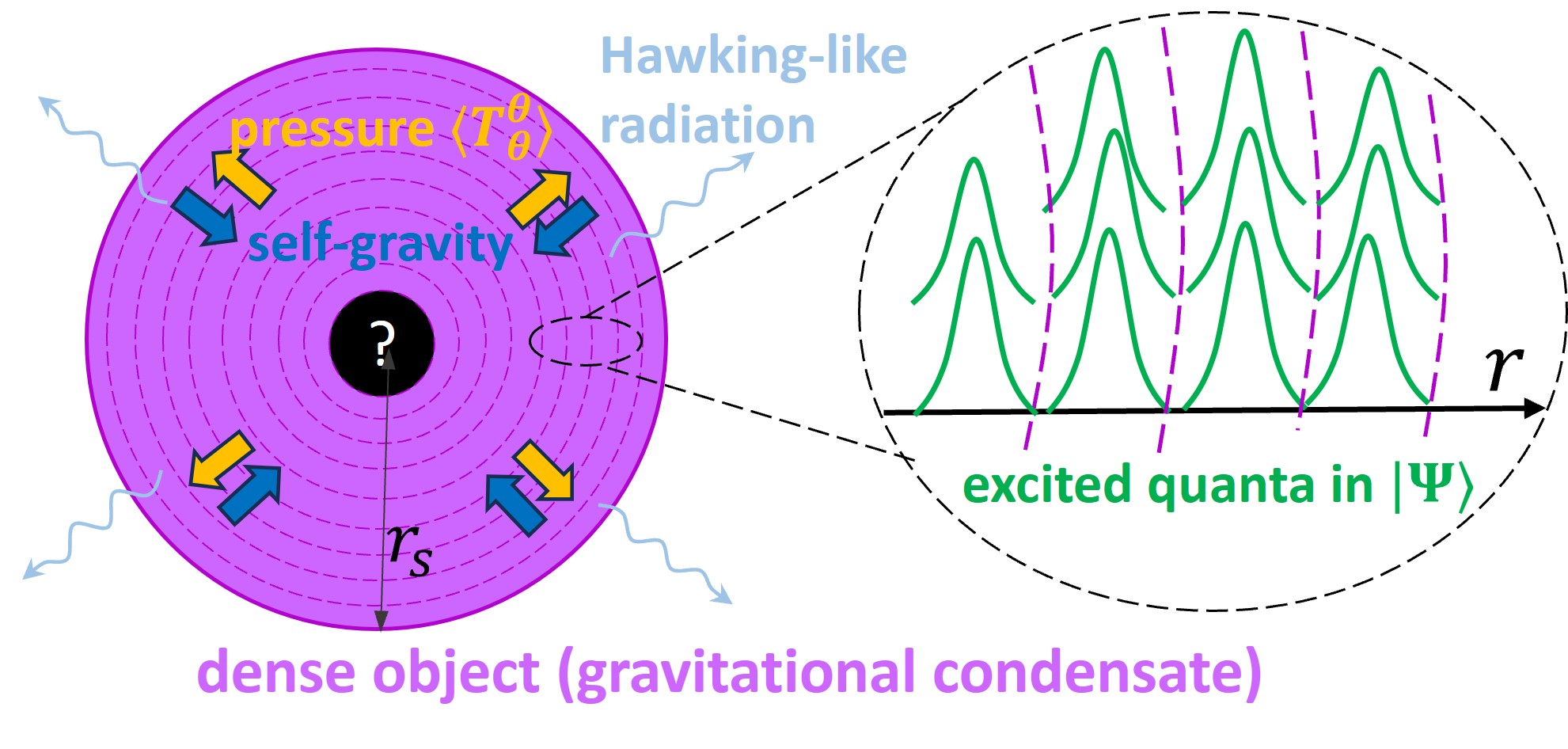}
\caption{The dense object, gravitational condensate, represented by the KY metric \eqref{metric_KY}. The central region beyond the semi-classical approximation \eqref{semi_Einstein} is unclear yet, but it has been checked at least that no large singularity exists.}
\label{f:KY}
\end{center}
\end{figure}

Let us start with a brief review of the KY metric \cite{KY1}:
%
%It is given by \cite{KY1} 
\begin{equation}
\label{metric_KY}
ds^2 = - \frac{\eta^2\s}{2r^2}e^{-\frac{\radius^2-r^2}{2\eta \s}}dt^2+\frac{r^2}{2\s}dr^2+r^2d\Omega^2
\,.
%ds^2 = - \frac{\eta^2\s}{2r^2}e^{-\frac{R^2-r^2}{2\eta \s}}dt^2+\frac{r^2}{2\s}dr^2+r^2d\Omega^2,
\end{equation}
This line element is valid in the radial coordinate range $\sqrt{\sigma}\lesssim r\leq \radius$.
This metric ansatz depends on three parameters: $\sigma$, $\eta$ and $\radius$. It describes the interior of a dense object, resembling a conventional black hole for an exterior observer, but non-perturbatively stabilized by the feedback of renormalized quantum fields on the geometry, as illustrated by fig.\ref{f:KY}.

The parameters $\sigma$ and $\eta$ encode the relevant properties of the quantum matter, respectively, the number of quanta $n\sim \s /l_p^2$ and the parameter of the linear equation of state \eqref{eta_eq} of the quantum fluid. While $\sigma$ sets a minimal length scale, the parameter $r_s$ gives the radial position of the star surface. There, as shown in \cite{Y1}, the KY metric connects to the exterior Schwarzschild metric \eqref{Sch} 
%\be
%\rd s^2
%=
%-\left(1-\frac{a_0}{r}\right)dt^2 + \left(1-\frac{a_0}{r}\right)^{-1}dr^2 + r^2 d\Omega^2
%\,,
%\ee
consistently with Israel's junction conditions. The surface position is then directly determined by the ADM energy ${a_0}/{2G}$,
\begin{equation}\label{surface}
\radius=a_0+\frac{\sigma\eta^2}{2a_0},
\end{equation}
at leading orders for large mass, $a_0 \gg l_p$.
Note that we write here $r_s$ for the {\it surface radius}, instead of the Schwarzschild radius.
Indeed, it must be underlined that $\radius>a_0$ and that there is no actual horizon\footnotemark{}.
\footnotetext{
Note that the proper length between $r=\radius$ and $r=a_0$ is of order $\sqrt{n}l_p\gg l_p$.
}
%\be
%1-\frac{a_0}{r_s}
%\sim
%\f{\eta^2\sigma}{2\radius^2}
%\propto
%\left(
%\f{l_P}{a_0}
%\right)^2
%\,.
%\ee
Phenomenologically, however, the exponentially large redshift of the KY metric \eqref{metric_KY} makes the imaging almost black and, seen from the exterior, the object behaves as a black hole in astrophysical cases with non-Planckian masses.
The small deviation from classical Schwarzschild black holes is consistent with current observations \cite{CY}.

%
%The size $R$, the position of the surface, can be determined, consist with Israel's junction condition, as 
%\begin{equation}\label{surface}
%    R=a_0+\frac{\sigma\eta^2}{2a_0},
%\end{equation}
%where $\frac{a_0}{2G}$ is the ADM energy, 
%and the interior metric \eqref{metric_KY} connects to the exterior Schwarzschild metric \eqref{Sch} there \cite{Y1}. Note that the proper length between $r=R$ and $r=a_0$ is $\sim \sqrt{n}l_p\gg l_p$, and no horizon exists. This is consistent with the Buchdahl limit \cite{Buchdahl} because of the large anisotropy in pressures. 
%

As for the positive constants $\eta$ and $\sigma$, they are determined by solving the semi-classical Einstein equation,
\begin{equation}\label{semi_Einstein}
G_{\mu\nu}=8\pi G\bra\Psi| T_{\mu\nu}|\Psi\ket
\,.
\end{equation}
%Indeed, we can solve it self-consistently and
Indeed, it was shown in \cite{KY4} that the KY metric \eqref{metric_KY} is a non-perturbative solution  in $\hbar$ for  $1\leq \eta <2$ and $\sigma=\MO(n l_p^2)$, where $n$ is the number of the degrees of freedom that contribute to the entropy of the system.
More precisely, for $n$ massless scalar fields minimally-coupled to gravity, we have \cite{KY4}:
\be
\sigma=\frac{n l_p^2}{120\pi \eta^2}
\,,
\ee
while the works \cite{KY1,KY3} computed for conformal matter fields:
\begin{equation}
\label{sigma}
\sigma=\frac{8\pi l_p^2 c_W}{3\eta^2}
\,,
\end{equation}
where $c_W$ is a coefficient in the 4D conformal anomaly (see \eqref{anomaly} in section \ref{sec:KY_reg} for more details) and plays a role of $n$. Here, the number of degrees of freedom is assumed to be large, $n\gg1$, such that the curvature is kept (much) smaller than the Planck scale: 
\begin{equation}\label{RRR}
\sqrt{R_{\mu\nu\a\b}R^{\mu\nu\a\b}},~\sqrt{R_{\mu\nu}R^{\mu\nu}},~R=\MO\l(\frac{1}{n l_p^2}\r)
\,,
\end{equation}
for $r\gg l_p$.
This ensures that we remain in the semi-classical regime and that the semi-classical Einstein equations are the appropriate approximation.

Physically, the KY metric represents a dense configuration of self-gravitating excited quanta \cite{Y1,Y2}, similar in principle to a gravitational condensate as in \cite{Dvali, Oriti}, as illustrated on fig.\ref{f:KY}.
%\footnote{See Refs.\cite{Y1,Y2} for relations to other gravity condensate models \cite{Dvali,Oriti}.} See Fig.\ref{f:KY}. 
The quanta together with vacuum fluctuations generate a energy-momentum tensor, consistent with the semi-classical Einstein equations,
\begin{align}
\label{EMT_KY}
\bra \Psi|\hat T^t{}_t|\Psi\ket
&=
-\frac{1}{8\pi G}\frac{1}{r^2}
\,,\\
\bra \Psi|\hat T^r{}_r|\Psi\ket
&=
\frac{1}{8\pi G}\frac{2-\eta}{\eta}\frac{1}{r^2}
\,,\nonumber\\
\bra \Psi|\hat T^\theta{}_\theta|\Psi\ket
&=
\frac{1}{8\pi G}\frac{1}{2\sigma \eta^2}
\,,\nonumber
\end{align}
at leading order in the radial coordinate $r\gg l_p$.
The time and radial components satisfy the linear barotropic equation of state $p_r=(2-\eta)\eta^{-1}\rho$, \eqref{eta_eq}.
The null energy condition holds, but the angular components of the stress-energy tensor  break the dominant energy condition, $\bra T^\theta{}_\theta \ket\gg\bra -T^t{}_t\ket\sim\bra T^r{}_r\ket$.
As a result, the interior is an anisotropic fluid, which allows for compact configurations beyond the Buchdahl limit \cite{Buchdahl:1959zz}.
The strong tangential pressure originates from vacuum fluctuations of the modes induced, with various angular momenta, by the
%curvature in a consistent way with the
4D Weyl anomaly \cite{KY3,KY4}.
The pressure and self-gravity balance each other, so that the matter quanta are not concentrated towards the center of the star but still uniformly distributed in the radial direction, in such a way that the curvature remains finite.

The energy of the central region $0\leq r \lesssim\sqrt{\s}$ is bounded by roughly the Planck mass $m_p \equiv \sqrt{\hbar/G}$.
%of at most $\MO(m_p)$
\cite{KY1,KY4,Y1}.
This remains beyond the domain of applicability of the semi-classical  Einstein equations \eqref{semi_Einstein}, and thus can not be precisely described by the KY metric \eqref{metric_KY}.
Nevertheless, the energy bound prevents large singularities in the center, such as those seen in classical black holes. This central region should ultimately be described by a more fundamental dynamics that transcends semi-classical dynamics. Using our effective Hamiltonian approach, we will describe a possible scenario extending the KY metric to this central region in section \ref{sec:KY_reg}.

%The size $R$, the position of the surface, can be determined, consist with Israel's junction condition, as 
%\begin{equation}\label{surface}
%    R=a_0+\frac{\sigma\eta^2}{2a_0},
%\end{equation}
%where $\frac{a_0}{2G}$ is the ADM energy, 
%and the interior metric \eqref{metric_KY} connects to the exterior Schwarzschild metric \eqref{Sch} there \cite{Y1}. Note that the proper length between $r=R$ and $r=a_0$ is $\sim \sqrt{n}l_p\gg l_p$, and no horizon exists. This is consistent with the Buchdahl limit \cite{Buchdahl} because of the large anisotropy in pressures. 

\medskip

The KY metric leads to a couple of interesting physical insights.
%provides to physically interesting results.
First, it can be obtained from several paths. The original way \cite{KMY} is to study the time evolution of a spherical collapsing matter including the backreaction from the particle creation in the time-dependent spacetime according to \eqref{semi_Einstein}. The final object at equilibrium in a heat bath is described by the KY metric\footnotemark{}
%\eqref{metric_KY}
\cite{KY1,KY2,KY3,KY4,KY5}.
%Note here that particle creation occurs generically in a time-dependent metric, independently of the existence of a horizon.
%
\footnotetext{
There exist similar constructions incorporating various aspects of QFT fluctuations in the structure of black holes.
A successful example are black stars that include the backreaction of vacuum polarization at every point inside the compact object (through a renormalized stress-energy tensor), creating an onion structure with Hawking radiation between each shell generating a negative pressure balancing out the gravitational pull \cite{Barcelo:2007yk,Visser:2008rtf,Barcelo:2009tpa}.
%to describe the interior of black holes using vacuum polarization,
The dynamical structure of KY black holes is very similar, and further incorporates non-perturbative 4D quantum fluctuations (beyond the vacuum state) through a detailed balance of the 4D anomaly. The Hawking radiation within the compact object is a mere semi-classical expression of those quantum fluctuations.
%
%A more thorough comparison of the black star proposal to the KY scenario and its associated formation/evaporation processes \cite{KMY} would certainly shed lights on the relevance of the various QFT effects on the internal structure of black holes. This is postponed to future investigation.
}
Another path is to identify the KY metric as the entropy-maximizing configuration for a given surface area $\MA$, saturating the Bousso bound \cite{Y2}. 

Second, we can evaluate the entropy density $s(r)$ of the quanta composing the object, and integrate it over the bulk volume. This reproduces the Bekenstein-Hawking formula exactly \cite{Y1}:
\begin{equation}\label{S}
    S=4\pi \int^{r_s}_{\sim \sqrt{\s}}dr \, r^2 \sqrt{g_{rr}(r)}s(r)=\frac{\MA}{4 l_p^2},
\end{equation}
where $\MA=4\pi r_s^2= 4\pi a_0^2+\MO(1)$. This is a result of the self-gravity.
%This should be a result of bulk interaction of matter quanta and gravity at a near-Planckian energy scale \cite{Y2}. 
Furthermore, interactions between matter quanta and Hawking radiation inside may lead to a possible scenario for information recovery \cite{KY2}.  

\medskip

In summary, the KY metric is a candidate for black holes consistent with quantum field physics and observation, grounded in a solid non-perturbative analysis of the semi-classical Einstein equations for geometry sourced by renormalized quantum matter.
%\footnote{Here, rather than defining black holes by the existence of a horizon, we are trying to describe objects recognized as black holes by observation in a manner consistent with quantum theory.} %It is a non-perturbative solution, and the energy scale is close to the Planck scale. This strongly suggest that there exists a more proper (potentially, simpler) description that connects semi-classical gravity and quantum gravity including matter fields. 
%
%Therefore, it should have a potential to connect semi-classical gravity and quantum gravity including matter fields. 
%
Therefore, it could be considered as a semi-classical template for quantum black hole physics, interfacing between the deep quantum gravity regime near the black hole center and the classical Schwarzschild regime describing the black hole exterior space-time.

Nevertheless, the framework can still be clarified and explored further.
In particular, understanding the essence of the self-consistent dynamics is a crucial challenge.
By ``self-consistent", we mean systematically fully taking into account of the non-linear feedback of the quantum field fluctuations on both the stress-energy tensor (through QFT renormalization) and the geometry (through the Einstein equation). This includes QFT effects on curved space-times, such as Hawking radiation. 
One method to show the self-consistency is to solve the Heisenberg equation of quantum matter fields in the KY metric and evaluate each component of the renormalized energy-momentum tensors \cite{KY4}, which is direct but complicated to implement. Another one is to apply the 4D Weyl anomaly in various ways \cite{KY3,KY5,HKLY}, which allows a clear non-perturbative analysis but requires extra assumptions. Both ways, the full 4D quantum fluctuation induced by the near-Planckian curvature makes the KY metric into a non-perturbative solution to the semi-classical Einstein equations.
%\sout{Previous methods involve evaluating the renormalized energy-momentum tensor, which is direct but complicated to implement (see e.g. \cite{KY4}), while the more recent approach consists in ensuring self-consistency through a computation of the 4D Weyl anomaly \cite{KY5,HKLY}. This makes the KY metric a non-perturbative solution to the semi-classical Einstein equations. This also}
%
This situation suggests that an appropriate description beyond the semi-classical Einstein equations, closer to quantum gravity, may exist. Such a description should be simpler yet more complete, capable of describing even the central region, and may advance our understanding of the general dynamics of quantum gravity.

In the following, we take a step in this direction, by investigating effective Hamiltonians consistent with the KY metric (in the next subsection) and searching for ones that could describe the whole black hole interior including the central region (in section \ref{sec:KY_reg}). This would give us further insight on modeling the feedback of quantum fluctuations of matter fields on the gravitational dynamics and on the mechanisms that allow to avoid space-time singularity.

%%%%%%%%%%%%%%%%%%%%%%%%%%%%%%%%%%%%%%
\subsubsection{Effective Hamiltonian for KY metric}
\label{sec:Heff_KY}
%%%%%%%%%%%%%%%%%%%%%%%%%%%%%%%%%%%%%%

Let us start by constructing a type $(+1)$ Hamiltonian, i.e. an effective energy density $\rhoeff[a,r]$, which leads to the KY metric \eqref{metric_KY},
\begin{equation}
\label{ab_KY}
a(r)=r-\frac{2\sigma}{r}
\,,\quad
b(r)=b_0 e^{\frac{r^2}{4\sigma \eta}}
\,,
\end{equation}
only for $l_p\ll r \leq r_s$,
where the constant factor $b_0=\frac{\eta}{2}e^{-\frac{\radius^2}{4\eta\s}}$ encodes the mass (and thus the surface radius) of the black hole. 
%Note again that this is valid for $l_p\ll r \leq R$. 
%
Such an effective Hamiltonian is supposed to reflect the full non-perturbative coupling of quantum matter fields on the geometry (such as the curvature generated by the one-loop renormalized  QFT corrections).

To construct a consistent effective energy density $\rhoeff(a,r)$, we consider the polynomial ansatz \eqref{Heff_P}  and follow the same logic as earlier, deducing $\rhoeff[a,r]$ from the equations of motion for the metric components $a$ and $b$. This leads to:
\begin{align}\label{rho_KY}
2G \rhoeff[a,r]=&\frac{r^2}{2\sigma \eta}+1-\frac{1}{\eta}+\frac{2\sigma}{r^2}+\frac{c_2}{2}\l(r-\frac{2\sigma}{r}\r)^2
\nonumber\\
&-\l[\frac{r}{2\sigma\eta}+c_2\l(r-\frac{2\sigma}{r}\r)\r]a+\frac{c_2}{2}a^2\,,
\end{align}
where $c_2$ remains an arbitrary function of the radial coordinate $r$. This provides a whole family of quadratic effective Hamiltonians consistent with the KY space-time structure.

For instance, choosing $c_2(r)=0$ simplifies this expression down to an effective Hamiltonian linear in $a$:
\begin{equation}
\label{rho_KY1}
2G \rhoeff=\frac{r^2}{2\sigma \eta}+1-\frac{1}{\eta}+\frac{2\sigma}{r^2}-\frac{r}{2\sigma\eta}a
\,.
\end{equation}
We can check that solving the corresponding equations of motion for $a$ and $b$, \eqref{eq_a} and \eqref{eq_b}, leads back to the desired behavior only for $r\gg l_p$:
\begin{equation}
    a(r)=r-\frac{2\sigma}{r}+C e^{-\frac{r^2}{4\sigma \eta}}\approx r-\frac{2\sigma}{r},~~b(r)=b_0 e^{\frac{r^2}{4\eta \sigma}},
\end{equation}
 where $C$ is a constant of integration.

Another possible choice is $c_2(r)=-\frac{1}{\sigma}$, leading to
\begin{align}
\label{rho_KY2}
2G\rhoeff
=&
-\l(1-\frac{1}{\eta}\r)\frac{r^2}{2\sigma}+3-\frac{1}{\eta}
\nn\\
&+\l[\l(2-\frac{1}{\eta}\r)\frac{r}{2\sigma}-\frac{2}{r}\r]a-\frac{a^2}{2\sigma}.
\end{align}
Solving exactly the equation of motion for $a$ leads to the desired behavior only for $r\gg l_p$,
\begin{align}\label{a_KY2}
a(r)&=r-\frac{2\s}{r}-\frac{C e^{-\frac{r^2}{4\eta\s}}}{\sqrt{\frac{\pi \eta}{4\s}}\l(\frac{2}{\sqrt{\pi}}+C \mathrm{erfc}(\frac{r}{\sqrt{4\eta\s}})\r)}
\nonumber\\
&\underset{r\gg l_p}\approx r-\frac{2\s}{r}
\end{align}
where $\mathrm{erfc}(x)\equiv1-\frac{2}{\sqrt{\pi}}\int^x_0dy e^{-y^2}$.
Thus, we once again reproduce the KY metric, as expected.
This illustrates the degeneracy in identifying effective Hamiltonians consistent with a given metric. Choosing a specific Hamiltonian over the other possibilities will have to be motivated by other considerations, such as a matching with modified gravity action principles or a direct derivation from a fundamental Lagrangian coupling renormalized matter fields to general relativity.\footnote{Again, we can select a special one $c_2(r)$ by asking the consistency with \eqref{Heff_c} in the linear level for $r\gg l_p$.}

We can go a step further and even identify consistent effective Hamiltonians beyond the quadratic truncation in $a$. For instance, we consider a cubic ansatz:
\begin{equation}\label{rho_1_3}
\rhoeff[a,r]=\frac{1}{2G}(c_1 a+\frac{c_3}{3}a^3)
\,.
\end{equation}
%using the metric \eqref{ab_KY} and solving \eqref{eq_a} and \eqref{eq_b} for $c_1$ and $c_3$,
Plugging this ansatz in the equations of motion for $a$ and $b$ gives, for $r\gg l_p$:
\begin{align}
\label{cc_KY}
c_1(r)&=\frac{3}{2r}\frac{r^2+2\sigma}{r^2-2\sigma}+\frac{r}{4\s \eta}\approx \frac{r}{4\sigma\eta}\nn\\
c_3(r)&=-\frac{3r}{2(r^2-2\sigma)^2}\l(\frac{r^2+2\sigma}{r^2-2\sigma}+\frac{r^2}{2\sigma \eta}\r)
\nn\\
&\approx-\frac{3}{4\sigma\eta r}\l(1+\frac{4\sigma}{r^2}\r)
\,.
\end{align}
Reversely, assuming those expressions, we can solve exactly the equations of motion for the space-time metric.
%In turn, we solve the equations of motion \eqref{eq_a} and \eqref{eq_b} with these coefficients.
Setting $\tilde{a}\equiv 1/a^2$, the equation of motion \eqref{eq_a} for the metric component $a$ reads:
%with \eqref{rho_1_3} becomes 
\begin{equation}
    \frac{d}{dr}\tilde{a}=-2c_1 \tilde a-\frac{2}{3}c_3, 
\end{equation}
and the solution is given by 
\begin{align}
\tilde a(r)=\l(C-\frac{2}{3}\int^r_{r_0}dr' c_3(r')e^{2\int^{r'}_{r_0}dr'' c_1(r'')} \r)e^{-2\int^r_{r_0}dr' c_1(r')},
\nn
\end{align}
where $C$ and $r_0$ are two integration constants\footnotemark{}.
%where $C$ is an integration constant and $r_0=\MO(\sqrt{n}l_p)$.
%
\footnotetext{
The extra integration constant $C$ does not affect the behavior of the metric at $r\gg l_P$, but it suggests various versions and completions of the KY metric to the central region $0\le r\lesssim l_p$.
}
Plugging the explicit expressions for $c_1$ and $c_3$ yields:
%Applying \eqref{cc_KY}, this becomes 
\begin{align}
    \tilde a(r)%&=\l(C+\frac{1}{2\sigma\eta}\int^r_{r_0}dr' \frac{1}{r'}\l(1+\frac{4}{r'^2}\r)e^{\frac{r'^2-r_0^2}{4\sigma\eta}} \r)e^{-\frac{r^2-r_0^2}{4\sigma\eta}}\nn\\
    &=Ce^{-\frac{r^2-r_0^2}{4\sigma\eta}}+\frac{1}{2\sigma\eta}\int^r_{r_0}dr' \frac{1}{r'}\l(1+\frac{4\sigma}{r'^2}\r)e^{-\frac{r^2-r'^2}{4\sigma\eta}}\nn\\
    &\approx Ce^{-\frac{r^2-r_0^2}{4\sigma\eta}}+\frac{1}{2\sigma\eta}\frac{1}{r}\l(1+\frac{4\sigma}{r^2}\r)\int^r_{r_0}dr' e^{-\frac{r(r-r')}{2\sigma\eta}}\nn\\
    &\approx Ce^{-\frac{r^2-r_0^2}{4\sigma\eta}}+\frac{1}{r^2}\l(1+\frac{4\sigma}{r^2}\r)\l(1-e^{-\frac{r(r-r_0)}{2\sigma\eta}}\r)\nn\\
    &\approx \frac{1}{r^2}\l(1+\frac{4\sigma}{r^2}\r),
\end{align}
where the approximations hold for $r\gg r_0\gg l_p$.
%
%where at the second line we have used $e^{-\frac{r^2-r'^2}{4\sigma\eta}}=e^{-\frac{(r+r')(r-r')}{4\sigma\eta}}\approx e^{-\frac{r(r-r')}{2\sigma\eta}}$ and picked up the most dominant contribution, and at the last line we have dropped exponentially small terms for $r\gg r_0$. Noting $\tilde a\equiv 1/a^2$,
%
This reproduces the desired mass function $a(r)$ for $r\gg l_p$. One can similarly check that we recover the desired lapse factor $b(r)$.
%from \eqref{eq_b} with \eqref{cc_KY}. 

This shows that our type-$(+1)$ ansatz for effective Hamiltonians, linear in the lapse $b$ and polynomial in the mass function $a$, is rich enough to contain a whole family of Hamiltonians generating the KY metric for $r\gg l_p$. Note that the KY metric does not solve these Hamiltonians exactly but approximately for $r\gg l_p$, thus
allowing flexibility for its completion in the Planckian region $0<r\lesssim l_p$.
More precisely, the expression of the effective energy density $\rhoeff(a,r)$ is not unique, and there are various expressions of $\rhoeff(a,r)$ that reproduce the same KY metric for $r\gg l_p$ and describe different Planck-scale physics for $0\leq r \lesssim l_p$. 
%On the one hand, having multiple effective Hamiltonian leading to the KY metric underlines a certain universality of this solution. On the other hand, it shows that certain (combinations of the) coefficients $c_n$ in the polynomial expansion of $\rhoeff$ are relevant to semi-classical behavior of the space-time metric (for $r\gg l_P$), while others are only relevant to the microscopic regime for the Planck scale physics at $0\leq r \lesssim l_p$.
%More precisely, the expression of the effective energy density $\rhoeff(a,r)$ is not unique, and there are various expressions of $\rhoeff(a,r)$ that reproduce the same KY metric for $r\gg l_p$.
%On the one hand, having multiple effective Hamiltonian leading to the KY metric underlines a certain universality of this solution. On the other hand, it shows that certain (combinations of the) coefficients $c_n$ in the polynomial expansion of $\rhoeff$ are relevant to semi-classical behavior of the space-time metric (for $r\gg l_P$), while others are only relevant to the microscopic regime for the Planck scale physics at $0\leq r \lesssim l_p$.

Then, one way to choose ``natural" Hamiltonians from various admissible candidates is to consider the completion of the KY space-time to the whole black hole interior. As shown in the next section, this will allow us to identify a ``simple" effective Hamiltonian that realizes a fully regular completion of the KY metric without singularity for the whole interior region $0\le r \le r_s$.
\section{Hamiltonian for Non-Singular Metrics}
%\section{Effective Hamiltonian for Non-Singular Metrics}
%\section{Effective Hamiltonian Avoiding Singularities}
%\section{Effective Hamiltonian for Regular Metrics}
\label{sec:regular}
%%%%%%%%%%%%%%%%%%%%%%%%%%%%%%%%%%%%%%%

%As another application, let us explore dynamics that can remove singularities using the effective Hamiltonian \eqref{Heff} with \eqref{rhoeff}. It is not obvious a priori whether the classical geometric description remains valid in the vicinity of $r=0$ of quantum black holes. We here assume this and construct the effective energy density $\rhoeff^{reg}(a,r)$ for non-singular metrics (Sec.\ref{sec:rho_reg}). We then derive a more complete version of KY metric (Sec.\ref{sec:KY_reg}).

Now that we have shown how to derive effective Hamiltonians reproducing known black hole metrics, we would like to push the logic further. Switching the focus back to the Hamiltonian as the more fundamental object to encode the dynamics of geometry, we would like to investigate further predictions of the formalism for the geometry in the vicinity of $r=0$ of (quantum) black holes.
First, we study, in section \ref{sec:rho_reg}, the conditions to impose on the effective energy density $\rhoeff[a,r]$ in order to generate non-singular metrics.
Then, in section\ref{sec:KY_reg}, we apply this formalism to the KY metric and explore effective Hamiltonians that generate regular completion of the KY metric with no singularity at the center $r=0$. We indeed provide a simpler energy density $\rhoeff^{reg}[a,r]$ that generates a metric valid for the whole black hole interior, with constant negative Ricci scalar, that fits the KY metric in the region $l_P\ll r \leq r_s$ and that connects with the standard Schwarzschild metric in the outside region $r\ge r_s$.

%%%%%%%%%%%%%%%%%%
\subsection{Conditions on the Energy Density}
%\subsection{Construction of $\rhoeff^{reg}(a,r)$}
\label{sec:rho_reg}
%%%%%%%%%%%%%%%%%%

We first examine the conditions on $a(r)$ and $b(r)$ for the spherically-symmetric static metric \eqref{metric} to be regular at $r=0$.
For simplicity, we here consider a ``typical" case for which 
\begin{align}\label{ab_r=0}
    \dot a \sim k r^n,~~\frac{d}{dr} \log b \sim k'r^m
\end{align}
for $r\to 0$, although it ignores possible logarithmic or exponential behaviors, which occur in some scenarios. 
Checking the behavior of the curvatures as $r\to0$, as proven in details in appendix \ref{A:regular}, one can show that the necessary behavior for the metric components are:
\begin{align}
\label{cond_a}
a(r) \underset{r\to0}{\propto} &
r^{n+1} &\textrm{with}\quad n\geq 2
\,,\\
\label{cond_b}
\frac{d}{dr} \log b(r) \underset{r\to0}{\propto} &
r^{m} &\textrm{with}\quad m\geq 1
\,,
\end{align}
or simply having vanishing $a$ and $\dot b$.
%
%\begin{align}
%\label{cond_a}
%    a(r)& \underset{r\to0}{\sim}
%    \left\{
%\begin{array}{ll}
%k r^{n+1} & (n\geq 2) \\
%{\rm or}\\
%{\rm const.}=0 
%\end{array}
%\right.,\\
%\label{cond_b}
%\frac{d}{dr} \log b(r)& \underset{r\to0}{\sim}
%    \left\{
%\begin{array}{ll}
%k' r^m & (m\geq 1) \\
%{\rm or}\\
%{\rm const.}=0  
%\end{array}
%\right.,
%\end{align}
%where $k,k'$ are constants.
%
Let us then see how these required behaviors translate to the effective energy density $\rhoeff[a,r]$. In particular, we would like to understand how to make the  the coefficients $c_k(r)$ consistent with those regularity conditions.

For simplicity, we assume a linear truncation of the energy density, that is $c_{k\ge 2}=0$.
%for $k\ge 2$.
It is rather natural to consider an energy density $\rhoeff=c_0 + c_1 a$ depending linearly in the mass $a$.
Moreover, as we have seen in section \ref{sec:Heffansatz}, this case is already rich enough to engineer which ever smooth metric components $(a(r),b(r))$.
%
%\begin{equation}\label{c_assump}
%    c_i(r)=0~~{\rm for}~~i\geq2.
%\end{equation}
%

Then, for such an energy density, the equation of motion of $b$ \eqref{eq_b} becomes $\rd_r\log b=-c_1$, which constrains the behavior of $c_1(r)$ for $r\to0$ to $c_1(r)\sim r^m$ with $m\geq1$.
Similarly, the equation of motion for $a$ \eqref{eq_a} implies that $c_0=\dot a -c_1 a \sim r^n$ for $r\to0$ with $n\geq 2$. Therefore, we obtain the effective Hamiltonian for regular metrics
$\Heff[a,b,r]=b(c_0(r)+c_1(r)a)/2G$ with:
%$\Heff[a,b,r]=b\rhoeff[a,r]$ with:
\be
\label{cond_c}
%2G\rhoeff[a,r]=c_0(r)+c_1(r)a
%\,,\quad
%\,\,\textrm{with}\quad
\left|
\begin{array}{cccc}
c_0(r) &\underset{r\to0}{\propto} &r^{n} &
\textrm{with}\,\, n\geq 2\,,
\vspace*{1mm}\\
c_1(r) &\underset{r\to0}{\propto} &r^{m} &
\textrm{with}\,\, m\geq 1\,,
\end{array}
\right.
\ee
with the possibility for $c_0$ or $c_1$ to simply vanish.
%\begin{align}
%\label{H_reg}
%\Heff^{reg}[a,b,r]&=b\rhoeff^{reg}[a,r]
%\\
%2G\rhoeff^{reg}[a,r]&=(c_0(r)+c_1(r)a)
%\nn
%\end{align} 
%with 
%\begin{align}
%\label{cond_c0}
%    c_0(r)& \underset{r\to0}{\sim}
%    \left\{
%\begin{array}{ll}
%\bar c_0 r^{n} & (n\geq 2) \\
%{\rm or}\\
%{\rm const.}=0 
%\end{array}
%\right.,\\
%\label{cond_c1}
%c_1(r)& \underset{r\to0}{\sim}
%    \left\{
%\begin{array}{ll}
%\bar c_1 r^m & (m\geq 1) \\
%{\rm or}\\
%{\rm const.}=0  
%\end{array}
%\right.,
%\end{align} 
%where $\bar c_0, \bar c_1$ are constants. 

%Here, we can find a natural interpretation of \eqref{rhoeff}. From \eqref{cond_a}, both $a(r)$ and $\dot a(r)$ must approach to zero as $r\to 0$. Also, $\dot a(r)$ is connected to $\rhoeff(a,r)$ through \eqref{Heq_II}. Therefore, a possible idea is that $\rhoeff(a,r)$ is expanded as a polynomial in a small $a$, which leads to the form of \eqref{rhoeff}. 

The simplest case of such a regular Hamiltonian is given by 
\begin{equation}
\rhoeff(r)=\frac{\bar c_0}{2G}r^2\,,
\end{equation}
with a constant $\bar c_0$.
By setting $\bar c_0=\Lambda$, the equation of motion for $a$ gives $a(r)=\frac{\Lambda}{3}r^3+C$. The regularity condition \eqref{cond_a} forces the integration constant to vanish, $C=0$. Similarly, the equation of motion for $b$ yields simply $b(r)=1$. We thus obtain the (anti-)de-Sitter metric, which is obviously fully regular, even at $r=0$.

%this agrees with the one for cosmological constant \eqref{Heff_c}. The equation of motion \eqref{eq_a} gives $a(r)=\frac{\Lambda}{3}r^3+C=\frac{\Lambda}{3}r^3$, where the condition \eqref{cond_a} required $C=0$, and \eqref{eq_b} derives $b(r)=1$. Therefore, the (anti-)de-Sitter metric \eqref{metric_dS} is obtained, which satisfies the conditions \eqref{cond_a} and \eqref{cond_b}.

%%%%%%%%%%%%%%%%%%%%%%%%%%%%%%
\subsection{Improved KY metric}
%\subsection{An improved KY metric}
\label{sec:KY_reg}
%%%%%%%%%%%%%%%%%%%%%%%%%%%%%%

A standard expectation is that non-perturbative quantum effects in general relativity naturally resolve the singularities at the heart of physical black holes. As KY metrics take into account the non-perturbative feedback of quantum matter fields on the geometry dynamics, it is therefore natural to expect a possible resolution of the black hole singularity. Following this logic, we would like to use the effective Hamiltonian framework to investigate possible regular extensions of the KY metric to the central zone of the black hole.

To this purpose, let us consider the simplest regular linear effective energy density,\footnote{We here focus on thermodynamically typical configurations and need $a$-dependence of $\Heff$ (from \eqref{rhoeff_rel}). Note that regular black-hole metrics in the literature \cite{1968qtr..conf...87B,Dymnikova:1992ux, Ayon-Beato:1998hmi, Dymnikova:2004zc, Hayward:2005gi} satisfy $\rhoeff+\peff = 0$ and are not typical.} satisfying the conditions discussed above.
As $c_0(r)$ is dimensionless and $c_1(r)$ has the dimension of the inverse of length, the smallest power laws satisfying the the conditions \eqref{cond_c} reads:
%Therefore, the conditions \eqref{cond_c} lead to the simplest one: 
\begin{equation}
\label{rho_reg}
2G\rhoeff^{reg}[a,r]=\frac{r^2}{k_0 l_p^2}-\frac{r}{k_0 k_1 l_p^2} a
\,,
\end{equation}
where we introduce Planck length factors to appropriately balance out the $r$-factors. The constant parameters $k_0,k_1$ are dimensionless such that $k_0 k_1 >0$ for the configuration to be typical (from \eqref{rhoeff_rel}). 
Comparing to the effective energy density \eqref{rho_KY1} for the KY metric which we derived in the previous section, we notice that this regular ansatz has the same structure except for a missing factor in $1/r^2$ and a constant term. We will further comment on this key difference below.

Solving the equations of motion for this new  effective Hamiltonian $\Heff^{reg}[a,b,r]=b\rhoeff^{reg}[a,r]$ yields:
\begin{align}
%\label{a_reg1}
a(r)&=k_1 r -l_p\sqrt{2k_0k_1^3}\cF\l(\frac{r}{l_p\sqrt{2k_0k_1}}\r)+C e^{-\frac{r^2}{2k_0k_1l_p^2}},
\nn\\
%\label{b_reg1}
\label{ab_reg1}
b(r)&=b_0 e^{\frac{r^2}{2k_0k_1l_p^2}},
\end{align}
where $\cF(x)\equiv e^{-x^2}\int^x_0 dy e^{y^2}$ is Dawson's integral, and $C$ is an integration constant. 
This lapse factor $b(r)$ directly agrees with the KY metric \eqref{ab_KY} upon matching the parameters $(k_0,k_1)$ with the KY parameters $(\sigma,\eta)$ such that $2k_0k_1l_P^2=4\sigma\eta$. This makes sense dimension-wise since $\eta$ is dimensionless and $\sigma$ has the dimension of an area. %Let's point out that $k_0k_1>0$ as long as we keep $\s>0$ and $\eta>0$ in the equation of state.
%
%For instance, choosing $\eta$

As for the mass function $a(r)$, we check its behavior close to $r=0$ and for large radial coordinate:
%We examine the asymptotic behaviors of \eqref{a_reg1}: 
\begin{align}
\label{a_reg_asym0}
a(r)
\underset{r\to 0}=
\frac{r^3}{3k_0l_p^2}+\MO(r^5)+C(1+\MO(r^2))
\,,
\end{align} 
\begin{align}
\label{a_reg_asym}
a(r)
\underset{r\gg l_p}=
k_1 r- \frac{k_0 k_1^2l_p^2}{r}+\MO(r^{-3})+C' e^{-\frac{r^2}{2k_0k_1l_p^2}}
\,,
\end{align} 
%
%\begin{align}
%\label{a_reg_asym}
%    a(r)&=
%    \left\{
%\begin{array}{ll}
%\frac{r^3}{3k_0l_p^2}+\MO(r^5)+C(1+\MO(r^2)) & {\rm{for}}~~r\to 0 \\
%k_1 r- \frac{k_0 k_1^2l_p^2}{r}+\MO(r^{-3})+C' e^{-\frac{r^2}{2k_0k_1l_p^2}} & {\rm{for}}~~r\gg l_p 
%\end{array}
%\right.,
%\end{align}
where $C'=C+i  l_p \sqrt{{\pi k_0 k_1^3}/{2}}$. % is a constant.
In the central region, for $r\to 0$, the regularity condition \eqref{cond_a} requires $C=0$.
In the semi-classical region, for $r\gg l_p$, the constant $C'$ is not relevant as long as $k_0k_1>0$, which is satisfied for typical fluids. 
%which is required by \eqref{rhoeff_rel} for typical excited configurations. 
Then the asymptotics of $a(r)$ fit exactly the KY metric \eqref{ab_KY} if we set $k_0l_p^2=2\sigma$ and $k_1=1$. This means imposing $\eta=1$ for the equation of state. 

\begin{figure}[h]
\begin{center}
\includegraphics*[scale=0.32]{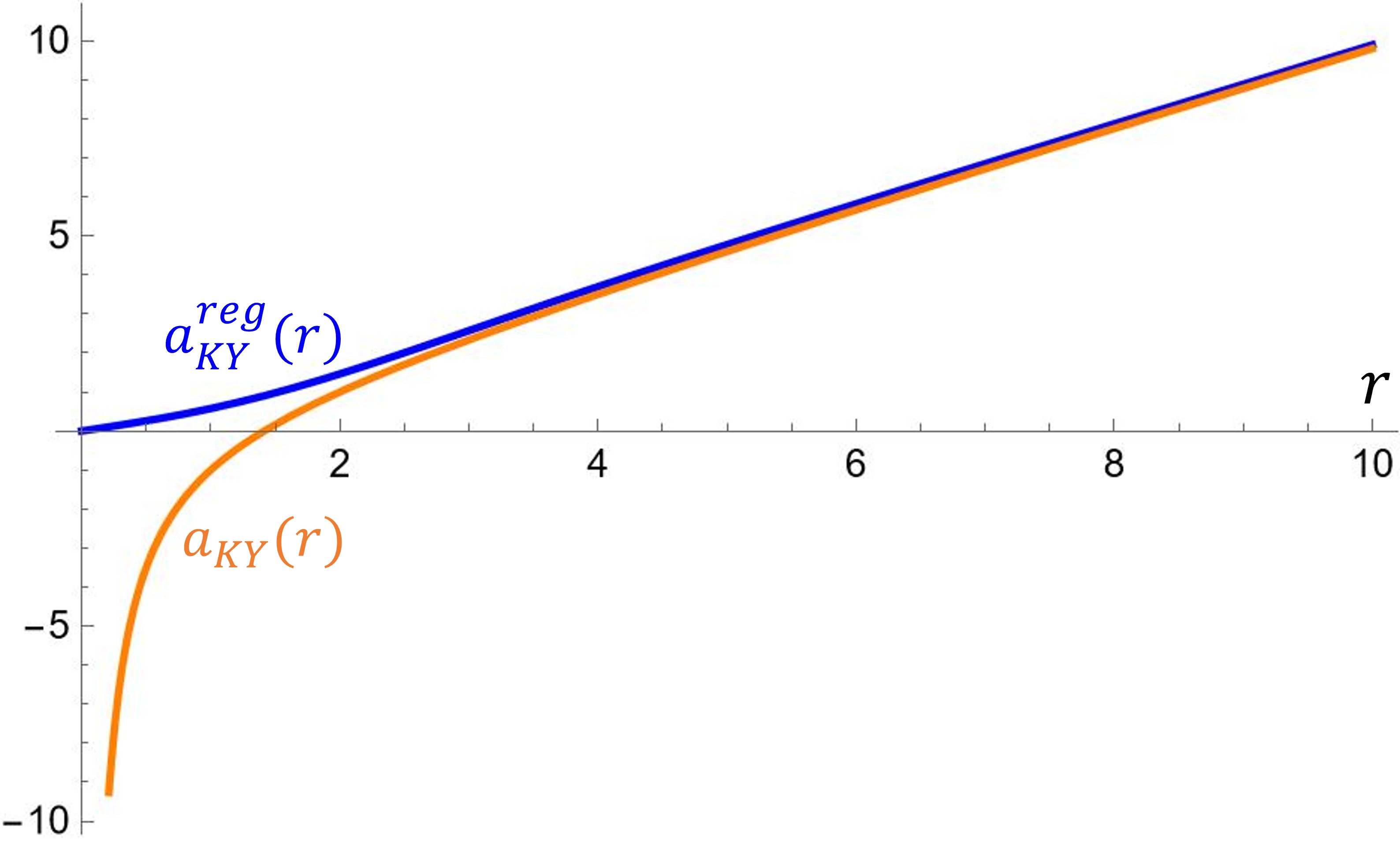}
\caption{Mass function $a(r)$ for $\sigma=1$:  \eqref{ab_KY_reg} (blue) and \eqref{ab_KY} (orange). Note that \eqref{ab_KY} is originally applicable only for $r\gg l_p$, and we here extrapolate it to $r=0$ formally.}
\label{f:aa_plot}
\end{center}
\end{figure}

At the end of the day, to summarize the analysis above, the effective energy density for this choice of parameters $(k_0,k_1)$ reads
\begin{equation}
\label{rho_reg_KY}
\rhoeff^{regKY}[a,r]
=
\frac{1}{2G}\l(\frac{r^2}{2\sigma }-\frac{r}{2\sigma} a\r)
\,.
\end{equation}
This generates the spherically-symmetric static metric with the following mass function and lapse factor:
\begin{align}
\label{ab_KY_reg}
a(r)=r -\sqrt{4\s}\cF\l(\frac{r}{\sqrt{4\s}}\r)
\,,\quad
b(r)=b_0 e^{\frac{r^2}{4\s}},
\end{align}
which we plot in figure \ref{f:aa_plot}.
The full metric reads:
\begin{align}
\label{metric_KY_reg}
\rd s^2
=&-\frac{\sqrt{4\s}}{4r}\cF\l(\frac{r}{\sqrt{4\s}}\r)e^{-\frac{r_s^2-r^2}{2\s}}dt^2
\nonumber\\
&+\frac{r}{\sqrt{4\s}\cF\l(\frac{r}{\sqrt{4\s}}\r)}dr^2+r^2d\Omega^2
\,.
\end{align}
In the central region, the mass function behaves at leading order as $a(r)\sim \frac{r^3}{6\s}$ for $r\to 0$, so that the metric looks like:
\begin{align}
%\label{metric_KY_reg}
\rd s^2
\underset{r\to 0}\sim&
%-\l(1-\frac{r^2}{6\sigma}\r)\frac{1}{4}e^{-\frac{r_s^2-r^2}{2\s}}dt^2
-\l(1+\frac{r^2}{3\sigma}\r)\frac{1}{4}e^{-\frac{r_s^2}{2\s}}dt^2
\nn\\
&+\l(1-\frac{r^2}{6\sigma}\r)^{-1}dr^2+r^2d\Omega^2
\,.
\end{align}
%\be
%\rd s^2
%=
%-\l(1-\frac{r^2}{6\sigma}\r)\frac{1}{4}e^{-\frac{r_s^2-r^2}{2\s}}dt^2
%+\l(1-\frac{r^2}{6\sigma}\r)^{-1}dr^2+r^2d\Omega^2
%\,.
%\ee
This metric is completely regular and does not have any singularity at $r\to 0$.
In the semi-classical region, the mass function behaves as $a(r)\sim r-2\s/r$ for $r\gg l_P$ leading to the asymptotic metric
\be\label{metric_KY_again}
ds^2\underset{r\gg l_P}\sim
-\frac{\s}{2r^2}e^{-\frac{r_s^2-r^2}{2\s}}dt^2+\frac{r^2}{2\sigma}dr^2+r^2d\Omega^2
\,,
\ee
which is exactly the KY metric for $\eta=1$.\footnote{The improved KY metric \eqref{metric_KY_reg} becomes approximately the original KY metric \eqref{metric_KY_again} for $r\gg l_p$. Therefore, following the argument of Ref.\cite{Y1}, the position of the surface satisfying the Israel junction condition can be identified as \eqref{surface}, as long as the total mass $\frac{a_0}{2G}$ is large. That is, for the Schwarzschild external metric and the original KY internal metric, requiring that all components of the surface energy-momentum tensor be as small as possible (i.e., that the interior and exterior be as continuous as possible) leads to the surface location being chosen as \eqref{surface}.}
%
%\begin{align}
%\label{a_reg_asym_KY}
%    a(r)&=
%    \left\{
%\begin{array}{ll}
%\frac{r^3}{6\s}+\MO(r^5)& {\rm{for}}~~r\to 0 \\
%r- \frac{2\s}{r}+\MO(r^{-3})& {\rm{for}}~~r\gg l_p 
%\end{array}
%\right..
%\end{align}
%%
%The metric is given by
%\begin{align}\label{metric_KY_reg}
%    ds^2&=-\frac{\sqrt{4\s}}{4r}F\l(\frac{r}{\sqrt{4\s}}\r)e^{-\frac{R^2-r^2}{2\s}}dt^2\nonumber\\
%    &~~~~~~~~~+\frac{r}{\sqrt{4\s}F\l(\frac{r}{\sqrt{4\s}}\r)}dr^2+r^2d\Omega^2,\\
%     &=\left\{
%\begin{array}{ll}\label{metric_KY_reg2}
%-\l(1-\frac{r^2}{6\sigma}\r)\frac{1}{4}e^{-\frac{R^2-r^2}{2\s}}dt^2\\
%    ~~~~~~~~~+\l(1-\frac{r^2}{6\sigma}\r)^{-1}dr^2+r^2d\Omega^2& {\rm{for}}~~r\to 0 \\
%-\frac{\s}{2r^2}e^{-\frac{R^2-r^2}{2\s}}dt^2+\frac{r^2}{2\sigma}dr^2+r^2d\Omega^2& {\rm{for}}~~r\gg l_p 
%\end{array}
%\right..
%\end{align}
%where we used $b_0|_{\eta=1}=\frac{1}{2}e^{-\frac{R^2}{4\s}}$ for the connection to the exterior Schwarzschild metric at $r=R$, \eqref{surface}.\footnote{
%In terms of $\bar t \equiv \frac{t}{\sqrt{4\s}}, \bar r\equiv \frac{r}{\sqrt{4\s}}$, \eqref{metric_KY_reg} can be expressed as 
%\begin{equation}
%    ds^2=4\s\l[-\frac{F(\bar r)}{4 \bar r}e^{-2(\bar R^2-\bar r^2)}d\bar t^2+\frac{\bar r}{F(\bar r)}d\bar r^2+\bar r^2 d\Omega^2
%    \r]\nonumber.
%\end{equation}}
%

That is, the metric \eqref{metric_KY_reg} describes the whole interior region of the black hole $0\leq r \leq r_s$. It fits with the KY metric with $\eta=1$ in the semi-classical region $r\gg l_p$ and defines a fully regular completion of that metric to the central region $r\to 0$. It can therefore be considered as an improved version of the KY metric.
Its dynamics contains the effects of quantum matter fields including quantum fluctuations consistent with the Weyl anomaly and excitations consistent with the entropy-area law, and possesses a mechanism that removes the singularity. It is remarkable that such a metric is derived naturally from the simplest effective Hamiltonian satisfying the regularity condition and the typical condition. This underlines the potential universality of this metric.

%%%%%%%%%%%%%%
\subsubsection{A simple effective Hamiltonian}
%%%%%%%%%%%%%%

We explore the properties of the effective Hamiltonian for the improved KY metric, which admits a surprisingly simple expression:
\begin{align}
\label{Heff_KYreg1}
    \Heff^{reg KY}&=\frac{b}{2G}\l(\frac{r^2}{2\sigma }-\frac{r}{2\sigma} a\r)\\
    \label{Heff_KYreg2}
    &=\frac{1}{2G}\frac{r^2}{2\s}bf
    \,,
\end{align}
where $b$ and $f$ are directly the factors entering our ansatz \eqref{eqn:THEMETRIC} for spherically-symmetric static metrics. The $r^2$ factor seems to be simply the area of the sphere resulting from the integration over the angular coordinates.

First, let us compare this expression to the effective Hamiltonian \eqref{rho_KY1} previously derived for the original KY metric with $\eta=1$: 
\begin{align}
\label{Heff_KY_eta1}
\Heff^{KY}|_{\eta=1}=\frac{b}{2G}\l(\frac{r^2}{2\sigma }+\frac{2\s}{r^2}-\frac{r}{2\sigma} a\r).
\end{align}
The only difference is the term ${2\s}/{r^2}$. Thus erasing this term allows to resolve the would-be singularity of the original KY metric  at $r=0$.
Nonetheless, this term $1/r^2$ can be somewhat intriguing.
Indeed, the combination $\frac{r^2}{2\sigma }+\frac{2\s}{r^2}$ in \eqref{Heff_KY_eta1} is invariant under the inversion map $r\to1/r$. This is reminiscent of UV-IR dualities appearing in both string theory \cite{String} and loop quantum gravity \cite{Modesto}.
%It would be interesting to investigate an effective action respecting such a duality in this framework.

%In the semi-classical region $r\gg l_p$, we 
%Note that $\Heff^{regKY}$ does not become $\Heff^{KY}|_{\eta=1}$ as $r\gg l_p$ while the metrics are related each other as in \eqref{metric_KY_reg2}, and that $\Heff^{KY}|_{\eta=1}$ is applicable for $r\gg l_p$.\footnote{The structure $\frac{r^2}{2\sigma }+\frac{2\s}{r^2}$ in \eqref{Heff_KY_eta1} is invariant under a transformation $r\to1/r$. This could be reminiscent of UV-IR dualities \cite{String, Modesto}. It would be interesting to investigate an effective action respecting such a duality in this framework.} This suggests that a transition at the dynamics level occurs between the central and surrounding parts (see below). 

Next, the $bf$ combination in this Hamiltonian $\Heff^{reg KY}$  is surprisingly similar to the effective Hamiltonian for the causal-limit fluid \eqref{Heff_causal}, which, we recall, is
\begin{equation}
\Heff^{causal}=\frac{1}{2G}bf
\,.\nn
\end{equation}
The only difference is the missing coordinate-dependent factor ${r^2}/{2\s}$. Since the causal-limit fluid metric still has a singularity at $r=0$ (as reviewed in section \ref{sec:causal}), it seems that this very simple $r^2$ factor plays a key role in removing the singularity and turning the causal-limit fluid into the gravity condensate.
Moreover, the simple $bf$ combination appearing in both cases seems to demand a deeper physical insight and more fundamental explanation, which we haven't yet identified.
%
%The entropy of the fluid is also proportional to the surface area (but smaller than the Bekenstein-Hawking entropy) \cite{Y2,Banks}. Note that $\Heff^{causal}$ cannot be applied to $r=0$; extrapolating the metric \eqref{metric_Zel} to $r=0$ would lead to a singularity. Therefore, the factor $\frac{r^2}{2\s}$ should play key roles in removing the singularity and changing the fluid into the gravity condensate.   

%Thus, we have two complementary approaches to understand the key of the dynamics resolving singularities. One is to consider a mechanism removing the term $\frac{2\s}{r^2}$ from $\Heff^{KY}$, and the other is to investigate how to add the factor $\frac{r^2}{2\sigma}$ to $\Heff^{causal}$. Although these may be model-dependent, they should provide clear and insightful perspectives on the singularity problem.

%%%%%%%%%%%%%%
\subsubsection{Stellar structure}
%%%%%%%%%%%%%%

We study the structure of the object described by this regular completion of the KY metric.
%the metric \eqref{metric_KY_reg}.
%
First, the Ricci scalar is constant in the whole interior, with a near-Planckian negative value:
\begin{equation}
\label{R_KY_reg}
R=-\frac{1}{\s}.
\end{equation}
This remarkable feature comes as a surprise. The original expectation was indeed to have a flat space core from the dynamical perspective of gravitational collapse \cite{KY3, KY4}. Moreover, singularity avoidance scenarios for black hole often involve negative pressure in the central region, creating a de Sitter-like geometry close to $r=0$ \cite{Dymnikova:1992ux,Dymnikova:2001fb,Mazur:2004fk,Mottola:2023jxl}.
In fact, if the factor $b(r)=b_0e^{\frac{r^2}{2\s}}$  were absent from the improved KY metric \eqref{metric_KY_reg}, the leading behavior of the  metric at the core would be the de Sitter metric  with $\Lambda=\frac{1}{2\sigma}$. However, this factor plays a crucial role. It makes the pressures positive and produces the entropy density responsible for the areal law \eqref{S}. 

Despite the Ricci scalar being of constant negative value, let us emphasize that the interior is not (equivalent to) an anti-de Sitter space. Indeed the Ricci tensor is not isotropic and corresponds to a non-trivial energy-momentum distribution: 
\be
\label{J_reg1}
-\cT{}_t{}^t=\cT{}_r{}^r=\frac{\cF\l(\frac{r}{\sqrt{4\s}}\r)}{8\pi G \sqrt{\s}~r}
=
%\nonumber\\=&
\left\{
\begin{array}{l}
\frac{1}{16\pi G \s}+\MO(r^{2})\, {\rm{for}}\,r\to 0\,,
\vspace*{2mm}\\
\frac{1}{8\pi Gr^2}+\MO(r^{-4})\, {\rm{for}}\,r\gg l_p\,, 
\end{array}
\right.
\nn
\ee
\begin{align}
\label{J_reg2}
&\cT{}_\theta{}^\theta=\frac{1}{16\pi G\s}
\,.
\end{align}

\begin{figure}[h]
\begin{center}
\includegraphics*[scale=0.4]{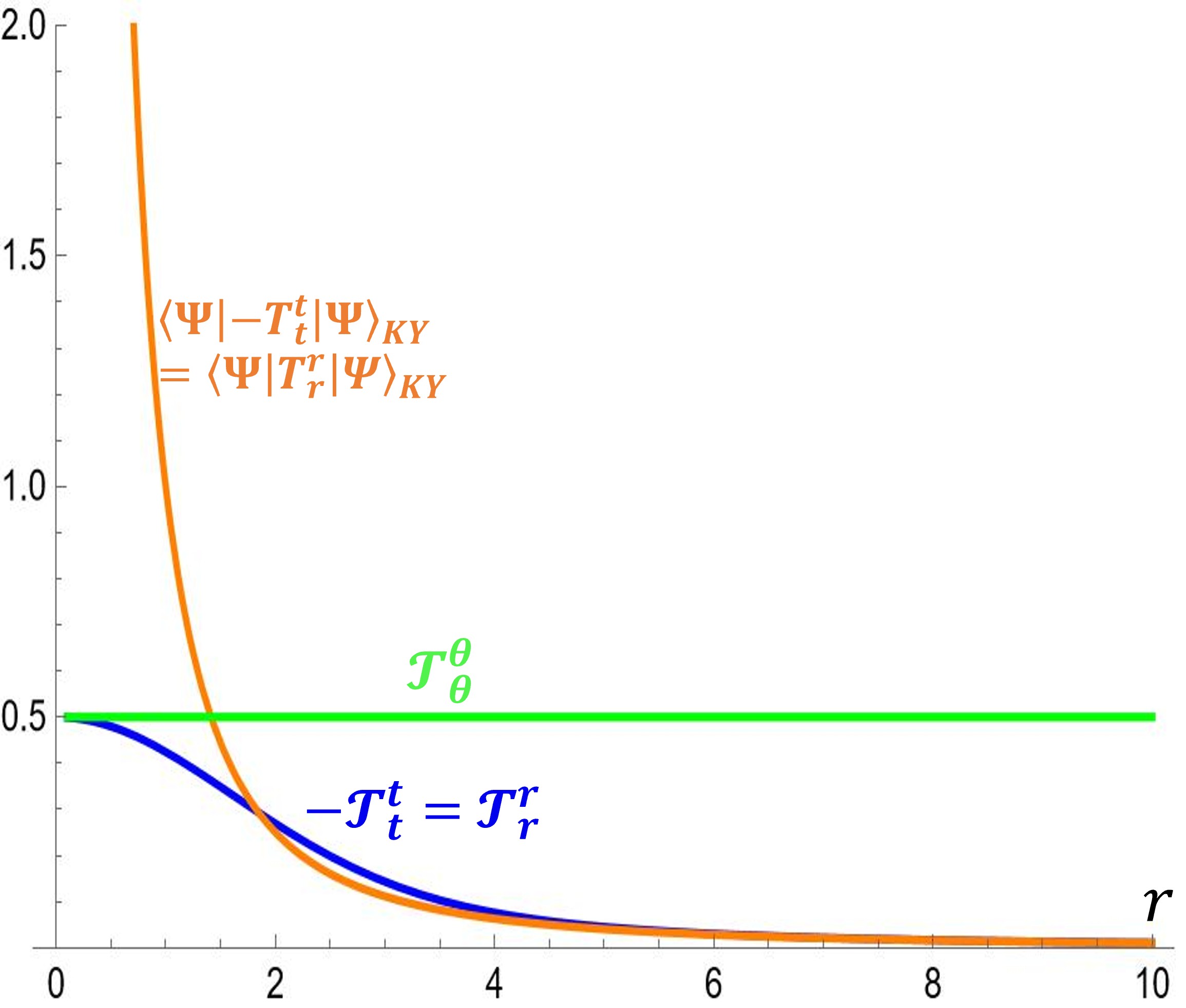}
\caption{$-\cT{}_t{}^t(r)=\cT{}_r{}^r(r)$ in \eqref{J_reg1}(blue), $\cT{}_\theta{}^\theta(r)$ in \eqref{J_reg2}(green), and (an extrapolation of) $\bra\Psi|-T{}_t{}^t(r)|\Psi\ket_{KY}=\bra\Psi|T{}_r{}^r(r)|\Psi\ket_{KY}|_{\eta=1}$ in \eqref{EMT_KY} (orange). $8\pi G=1$, $\sigma=1$.}
\label{f:TT_plot}
\end{center}
\end{figure}

As one can see on figure \ref{f:TT_plot}, this corresponds for $r\to0$ to a causal-limit isotropic fluid in the central region, with
\be
\frac{1}{16\pi G\s}
=\cT{}_\theta{}^\theta
=\cT{}_r{}^r+\MO(r^{2})
=-\cT{}_t{}^t+\MO(r^{2})
\,,
\ee
while it behaves like an anisotropic fluid in the semi-classical region for $r\gg l_p$, as expected for the original KY metric.

\begin{figure}[h]
\begin{center}
\includegraphics*[scale=0.7]{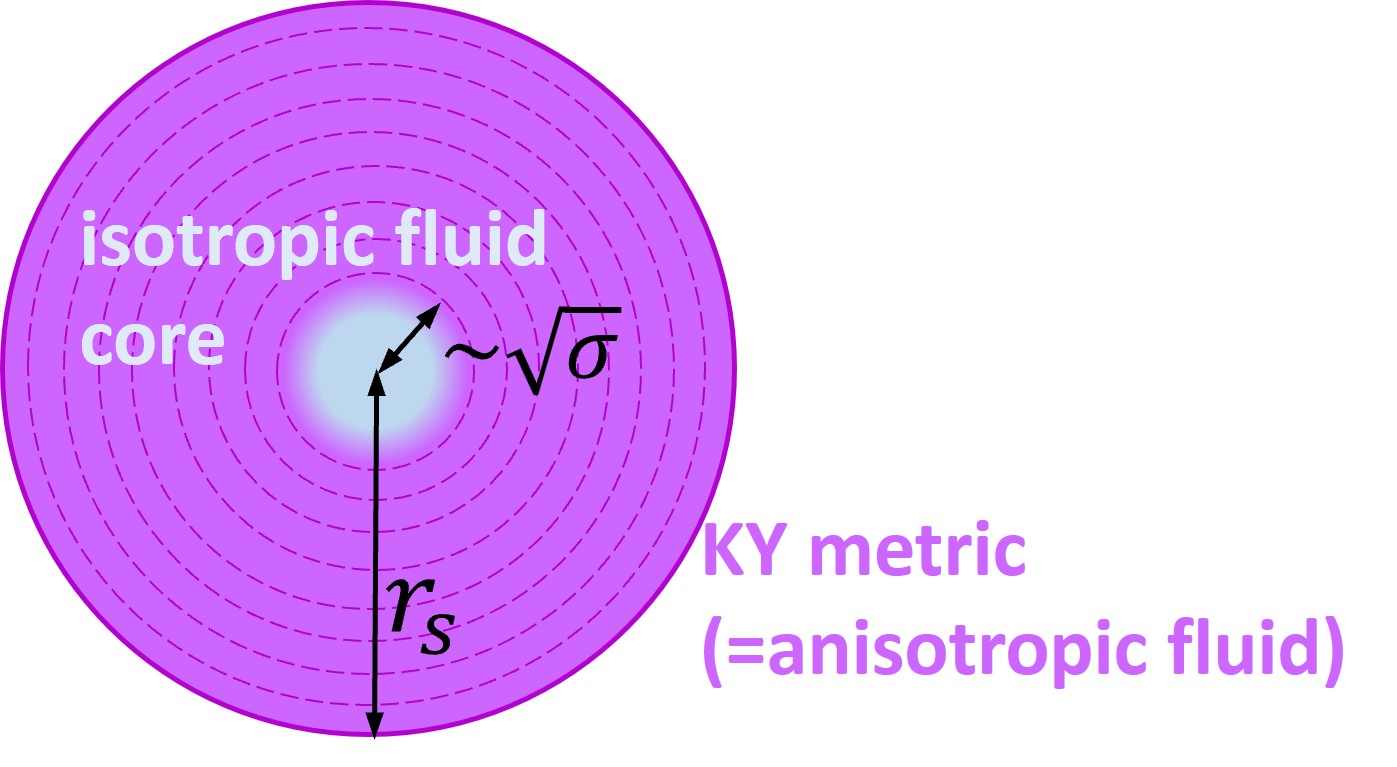}
\caption{The interior structure of the quantum black hole described by the improved KY metric \eqref{metric_KY_reg}.}
\label{f:KYreg}
\end{center}
\end{figure}

Therefore, as illustrated by fig.\ref{f:KYreg}, the black hole interior structure defined by the improved/regularized KY metric consists of an isotropic-fluid core covered by anisotropic fluid, until we reach the star surface, where the metric transitions to the standard Schwazschild metric.
%the center region $0\leq r \lesssim \sqrt{\s}$ is the isotropic-fluid core, and the intermediate region $\sqrt{\s}\lesssim r \leq R$ is the anisotropic fluid given by the original KY metric. 
%

%%%%%%%%%%%%%%
%\subsubsection{Transition in Dynamics}
%%%%%%%%%%%%%%

\medskip

Let us dive deeper into the difference of the dynamics between the (regularized) core part and the surrounding KY region.
%
%To model regular black holes, one often introduces a de-Sitter space inside \cite{XX}. 
%%
%Indeed, if the factor $b(r)=b_0e^{\frac{r^2}{2\s}}$ in the center metric of \eqref{metric_KY_reg2} were absent, it would be the de-Sitter metric \eqref{metric_dS} with $\Lambda=\frac{1}{2\sigma}$. However, the factor plays a crucial role; it makes the pressures positive and produces the entropy density, deriving the areal law \eqref{S} from the bulk. Therefore, the center metric could be considered as an excited version of the de-Sitter space, although the Ricci scalar is negative as in \eqref{R_KY_reg}. Note also that originally, the center region was expected to be flat space from a perspective of a dynamical model \cite{KY3, KY4}, while the improved KY metric suggests the AdS-like space \eqref{metric_KY_reg2}.
%
The change of behavior between isotropic and anisotropic fluids suggests a phase transition at the dynamics level. We show that this is reflected by a different value of the central charge in the two regions.

%To examine the validity of semi-classical dynamics, let us consider a simple analysis.
More precisely, let's assume that the dynamics in the whole interior is determined by the semi-classical Einstein equation
%\eqref{semi_Einstein}
coupled with conformal matter fields. Then, the trace part of the energy-momentum tensor is given by the 4D Weyl anomaly \cite{Birrell:1982ix,Duff}: 
\begin{equation}\label{anomaly}
    \bra \Psi|\hat T^\mu{}_\mu|\Psi\ket = \hbar (c_W C_{\a\b\mu\nu}C^{\a\b\mu\nu}-a_W\mathcal{G}+b_W \Box R), 
\end{equation}
\be
\textrm{with}\,\,
\left|
\begin{array}{l}
C_{\a\b\mu\nu}C^{\a\b\mu\nu}=R_{\a\b\mu\nu}R^{\a\b\mu\nu}-2R_{\a\b}R^{\a\b}+\frac{1}{3}R^2\,,
\vspace*{2mm}\\
\mathcal{G}=R_{\a\b\mu\nu}R^{\a\b\mu\nu}-4R_{\a\b}R^{\a\b}+R^2
\,,
\end{array}
\right.
\nn
\ee
%where $C_{\a\b\mu\nu}C^{\a\b\mu\nu}=R_{\a\b\mu\nu}R^{\a\b\mu\nu}-2R_{\a\b}R^{\a\b}+\frac{1}{3}R^2$ and $\mathcal{G}=R_{\a\b\mu\nu}R^{\a\b\mu\nu}-4R_{\a\b}R^{\a\b}+R^2$.
%
where $c_W$ and $a_W$ are positive constants fixed by the matter content of the theory,
%for small coupling constants,
while $b_W$ is a positive/negative constant depending on the couplings of the higher derivative $R^2$ and $R_{\a\b}R^{\a\b}$ (counter-)terms in the gravity action. 

We apply the formula \eqref{anomaly} to $G^\mu{}_\mu=8\pi G \bra \Psi|T^\mu{}_\mu|\Psi\ket$, to determine the value of the parameter $\s$ relfecting the number of degrees of quanta entering the regularized KY metric \eqref{metric_KY_reg}.
For $r\gg l_p$, we have 
\begin{align}\label{sigma_IR}
    \frac{1}{\s}=\frac{8\pi l_p^2c_W}{3\s^2}+\MO(r^{-2})\Rightarrow \s=\frac{8\pi l_p^2 c_W}{3}\equiv \s_{IR},
\end{align}
which agrees with the value \eqref{sigma} of KY metric for $\eta=1$. 
For $r\to0$, on the other hand, we have 
\begin{align}\label{trace_eq_r=0}
    \frac{1}{\s}=\frac{32\pi l_p^2a_W}{3\s^2}+\MO(r^{2})\Rightarrow \s=\frac{32\pi l_p^2 a_W}{3}\equiv \s_{UV},
\end{align}
which is different from $\s_{IR}$.
Since the parameter $\s$ is assumed to be constant in the metric \eqref{metric_KY_reg}, having $\s_{IR}\neq \s_{UV}$ is a clear contradiction\footnotemark{}, which highlights that the semi-classical Einstein equations are not valid in the whole interior. 
\footnotetext{One might think of the possibility of a  matter content satisfying $\s_{IR}=\s_{UV}$, thus with $c_W=4a_W$. However, this is not consistent with a constraint derived in \cite{Duff,Hofman}: 
$\frac{1}{3}\leq \frac{a_W}{c_W}\leq\frac{31}{18}$.} 

As expected, this underlines that the semi-classical Einstein equation is only applicable to the  $\sqrt{\s}\lesssim r \leq R$ region, and that a clean physical description of the center core requires a new quantum gravity dynamics beyond the semi-classical treatment, which would lead to a revised value of $\s_{UV}$ matching  $\s_{IR}$. 

%Consistency with the effective Hamiltonian \eqref{rho_reg_KY} that encompasses both dynamics should provide insight into the microscopic degrees of freedom and dynamics that unify the descriptions of matter and gravity. In summary, the improved KY metric 

%%%%%%%%%%%
\subsubsection{Another scenario?}
The energy scale of the interior of the object described by the improved KY metric is close to the Planck scale, and it should be natural to assume that matter fields are massless and approximately conformal. As shown above, this led us to the mismatch of the two values of $\sigma$, meaning that the semi-classical Einstein equation is not valid at the central part. 

Nevertheless, let us here try to suggest another scenario where the semi-classical dynamics is kept. Suppose that $N$ new massive scalar fields $\phi_i$ with
Planck scale masses $m_i=\MO(m_p)$ appear around $r=0$ by some new physics (e.g. \cite{String}). Then, the right hand side in the trace part of the semi-classical Einstein equation \eqref{trace_eq_r=0} would gain additional terms $- 8\pi G \frac{1}{2}g^\mu{}_\mu \sum_{i=1}^Nm_i^2 \bra \phi_i^2 \ket $: 
\begin{align}
    \frac{1}{\s}&=\frac{32\pi l_p^2a_W}{3\s^2}-16\pi G N m_p^2 C\nn\\
    \Rightarrow NC&=\frac{3}{128\pi^2 c_W G}\l(\frac{4a_W}{c_W}-1\r),
\end{align}
where we assume $\bra \phi_i^2\ket \approx C={\rm const.}$ for $r\sim 0$ and set $\sigma=\sigma_{IR}$ \eqref{sigma_IR}. This would be another scenario within the semi-classical dynamics as long as the self-consistency is checked. 

At the end of the day, more work on new dynamics around the center would be required to settle this issue. 

%%%%%%%%%
\subsubsection{Self-interaction and the value of $\eta$.} 
%%%%%%%%%%

Assuming a regular metric and an energy density ansatz\eqref{rho_reg} linear in $a$, we got the effective Hamiltonian \eqref{rho_reg_KY} and derived a regular KY metric. But this worked only for an equation of state with $\eta=1$.
%
%Under the assumption \eqref{c_assump}, we have obtained the effective Hamiltonian for regularity \eqref{H_reg}, and the simplest non-trivial one \eqref{Heff_KYreg1} has led to the improved KY metric \eqref{metric_KY_reg}. Here, only $\eta=1$ is allowed for the connection to the original KY metric in $r\gg l_p$.
%
So, is it possible to obtain other values of the parameter $\eta$?

A natural idea is to move beyond the linear truncation of the energy density and include higher terms in $a$. These would represent higher order self-interaction terms in the mass $m(r)\equiv\frac{a(r)}{2G}$ via $G$. This would be consistent with the original motivation for $\eta\neq1$  explained in \cite{KY1}: the effect of interactions between radiation and quanta that constitute the compact object is represented phenomenologically by various values of the parameter $\eta\neq1$ entering the equation of state \eqref{eta_eq}.

To start with, let us study a small deviation from $\eta=1$ by a perturbation parameter $\xi\equiv \frac{\eta-1}{\eta}\ll1$ and see if it can be obtained by adding a quadratic term in the energy density. We add a self-interaction term to  \eqref{Heff_KYreg1}: 
\begin{equation}\label{rho_KY_int}
    \Heff^{regKYint}(a,r)=\frac{1}{2G}\l(\frac{r^2}{2\sigma }-\frac{r}{2\sigma} a + \frac{k\xi}{2\s}a^2\r),
\end{equation}
%where we used the fact that $c_2(r)$ has the dimension of the inverse of the square of the length,
where we look for a suitable coupling constant $k$.\footnote{Note here that all terms in the bracket are dimensionless and should be proportional to $\frac{1}{\sigma}\sim \frac{1}{l_p^2}$, as expected as a non-perturbative effect. Therefore, the coefficient function of the quadratic term should be independent of $r$.} In order to solve for $a(r)$   and $b(r)$ perturbatively, we expand the solutions to the equations of motion as:
\begin{align}
    a(r)=a_*(r)+\Delta a(r),~\beta(r)=\beta_*(r)+\Delta\beta(r),
\end{align}
where we introduced $\beta\equiv \log b$. The non-perturbative part $(a_*,\beta_*)$ was derived previously in \eqref{ab_KY_reg}, while the perturbation $(\Delta a,\Delta \beta)$ is assumed to be of order  $\MO(\xi)$. The equation of motion for $b$ becomes at linear order in $\xi$:
\begin{align}
    &\frac{d}{dr} \Delta \b=-\frac{k\xi}{\s}a_*\approx -\frac{k\xi}{\s}r~~{\rm{for}}~~r\gg l_p\nonumber\\
    &\Rightarrow \Delta \beta=-\frac{k\xi}{2\s}r^2.
\end{align}
Therefore, if we choose $k=\frac{1}{2}$, we can fit the expected behavior for arbitrary value $\eta$. Indeed, we obtain (for $r\gg l_p$) 
\begin{equation}
    \b=\b_*+\D \b=\frac{r^2}{4\s}-\frac{\xi r^2}{4\s}=\frac{r^2}{4\s\eta},
\end{equation}
which agrees with the KY metric \eqref{ab_KY} for $\eta\neq 1$. Then, the equation of motion for $a$ reads
\begin{align}
    &\frac{d}{dr} \Delta a=-\frac{r}{2\s}\Delta a+\frac{\xi}{4\s}a_*^2\approx-\frac{r}{2\s}\Delta a+\frac{\xi}{4\s}r^2~~{\rm{for}}~~r\gg l_p\nonumber\\
    &\Rightarrow \Delta a=\frac{1}{2}\xi r- \frac{\xi\s}{r}+e^{-\frac{r^2}{4\s}}C+\MO(r^{-3})\approx \frac{1}{2}\xi r- \frac{\xi\s}{r}.
    \nn
\end{align}
We thus obtain,
\begin{equation}
    a=a_*+\Delta a=(1+\xi/2)\l(r-\frac{2\s}{r}\r)\,,
\end{equation}
which fits with our target KY metric \eqref{ab_KY}, unfortunately, only when $\xi=0$, i.e. $\eta=1$. 

This analysis hints that the value $\eta=1$ seems to be special and stable under perturbation of the effective Hamiltonian. Either this means that the equation of state $p_r=\rho$ has a more fundamental, yet to be understood, origin in this black hole scenario; or that one could reach arbitrary values of the parameter $\eta\neq1$ by a non-perturbative mechanism involving higher order self-interactions and thereby requiring going beyond the quadratic truncation of the effective Hamiltonian investigated above.

%%%%%%%%%%%%%%%%%%%%%%%%%%%%%%%%%%%%%%%
\section{Conclusion \& Outlook}\label{sec:dis}
%%%%%%%%%%%%%%%%%%%%%%%%%%%%%%%%%%%%%%%
%{\bf Points to include here
%\begin{itemize}
%\item theoretical aspects of improved KY ? solve dynamics including quantization \dots
%\item phenomenlogical aspects of improved KY ? POssibility of observaitonal signature of a constant negative bulk Ricci scalar?
%\item technical improvement of method towards linking with current astrophysical observations, by generalization to rotating and modified Kerr solutions, have to extend phase space for geometrical sector.
%\item How to lift the degeneracy for Hamiltonians? More investigation of derivation from fundamental action for gravity plus quantum matter. But also, understand link with modified gravity, e.g. Lagrangian with higher order terms. Put our effective approach in perspective of renromalization of general relativity.
%
%\item What is the symmetry for each effective Hamiltonian? Fluids have translation symmetry for thermal time? Or, can we use a symmetry to restrict the form of the Hamiltonian? 
%\end{itemize}
%}

%%%%%%%%%%%%%%%%%

As a first arena to investigate the dynamics of general relativity, the non-linear coupling of matter fields to the space-time geometry, and the relevance of modified gravity scenarios, spherically-symmetric static space-times are a non-trivial class of metrics exploring the physics of non-rotating black holes around and beyond the Schwarzschild metric.
We consider this reduced gravitational system (or "mini-superspace") as the leading order geometrical degrees of freedom of stellar structure.

In this context, the dynamics of spherically-symmetric static space-times can be formulated as the one-dimensional mechanics of the canonical pair of the Misner-Sharp mass $\frac{a(r)}{2G}$ and the lapse $b(r)$. The system is equipped with a Hamiltonian, encoding their evolution in terms of the radial coordinate $r$. In pure general relativity, without matter or modified gravity, this Hamiltonian simply vanishes. But in general, the coupling of geometry to classical or quantum matter, or the (quantum) fluctuations of the metric itself, will generate a non-vanishing effective Hamiltonian $\Heff[a,b,r]$. 

To support this approach, we explicitly derived the effective Hamiltonian driving the radial evolution of the geometry when coupled to a classical scalar field or to classical electromagnetism, showing that the induced dynamics lead back to the expected Janis-Newman-Winicour and Reissner-Nordstrom solutions, respectively.
Then we proposed a basic Hamiltonian ansatz, linear in the lapse factor $b$ and polynomial in the mass function $a$. The various powers in $a$ can be interpreted physically as self-interactions of the geometry, thus reflecting the non-linearity of the dynamics of the gravitational field in general relativity and modified gravity theories.

We illustrated the wide range of applicability of the approach by identifying effective Hamiltonians encoding the feedback of conformal fluids and causal-limit fluids on the geometry. We further applied the method to identify an effective Hamiltonian $\Heff^{KY}$ representing the dynamics of the Kawai-Yokokura (KY) solution of the semi-classical Einstein equations for geometry coupled to renormalized matter fields. It encodes non-perturbative quantum matter effects, consistent with the 4D Weyl anomaly and the entropy-area law, into the gravity dynamics. This shows that classical effective Hamiltonians for spherically-symmetric static space-times can also represent non-perturbative quantum physics.

Pushing further along a quantum gravity perspective, we discussed the condition to impose to the effective Hamiltonian in order to generate fully regular black hole metrics. Our main result is that the simplest such Hamiltonian ansatz is linear in both $a$ and $b$, and produces a regular version of the KY metric. This improved version of the KY solution represents a spherically symmetric dense region with the negative constant Ricci scalar. The black hole structure has a Planck-size singularity-free core, surrounded by a semi-classical region described by the original KY metric, and finally glued to the standard Schwarzschild metric at a surface of radial coordinate slightly larger than the Schwarzschild radius. The metric has no singularity and no horizon per se. The gravitational dynamics is stabilized by the quantum fluctuations of a fluid with linear barotropic equation of state. Generating such a solution from the simplest regular Hamiltonian shows the power of the approach.

\medskip

To push the physical relevance of the approach, we see two directions of investigation.
First, one could explore the phenomenology of the improved KY metric. Although the exponentially large redshift makes the imaging extremely similar to a standard black hole \cite{CY}, its gravitational wave spectrum could be significantly different and reveal non-perturbative effects of the coupling of renormalized quantum matter to gravity \cite{Cardoso:2019rvt}.
Second, one could generalize the method to rotating space-times. One would study the reduced phase space for a class of cylindrical metrics, e.g. the axisymmetric Weyl metrics \cite{Barrientos:2025abs}, and investigate the basic structure of effective Hamiltonians for this enlarged context. One could then seek a Kerr version of the KY solution. Incorporating rotation is a necessary step towards matching the theory to observed black holes.
%
%discuss a thermodynamical argument as a rotating system, consistent with the Weyl anomaly \cite{KY1}.

To set stronger foundations for our approach, one could understand how effective Hamiltonians fit with modified gravity, especially with higher curvature terms.
For instance, one could investigate if the (improved) KY black holes can be obtained from a higher derivative Lagrangian (e.g. general relativity with $R^2$ and $R_{\mu\nu}R^{\mu\nu}$ counter-terms). The higher order terms would account for the feedback of the renormalized quantum fields on the geometrical sector. This would reveal modified gravity theories with non-singular black hole solutions.
One could also look for systematic ways to derive effective Hamiltonians for wide classes of black holes obtained in modified gravity (e.g. \cite{BenAchour:2024hbg}).

Another line of investigation would be to extend our formalism to allow for time dependence as well as radial dependence.
One could use the so-called ``rigging technique" \cite{Rigging} in the $r$-foliation and construct a covariant version of the reduced gravity action used here \cite{PY}.
%, which has a similar structure to \eqref{Sg} even though the time integral $\int dt$ is included \cite{PY}.
The canonical variables would then consist of the Misner-Sharp mass, a gravitational pressure, the surface area of spatial sphere, and a 1-form that represents the volume form of the 2D spacetime normal to the sphere. Therefore, this would render possible to study a time-dependent version of the effective dynamics.
%This will be presented in upcoming work 

Finally, a classical Hamiltonian is the natural starting point for a quantization. We could study the (Schrödinger) quantization of our class of Hamiltonians, especially the ones generating the regular KY solutions, and analyze the magnitude of the resulting quantum fluctuations. We would check consistency with our semi-classical understanding of the KY black holes, but also explore which Hamiltonians generate large quantum fluctuations in the core region or close to the Schwarzschold radius, in order to sharpen the constraints to be satisfied by physical effective Hamiltonians.
This would be a systematic approach to clarifying the "mini-superspace" quantization of spherically-symmetric static space-time and classifying their behavior accordingly to basic features of their underlying Hamiltonian.

%%%%%%%%%%%%%%%%%

\bigskip

\stoptoc
\section*{Acknowledgments}
\resumetoc

E.L. acknowledges support from the CNRS
%International Research Project
IRP Grant "Quantum Gravty - Quantum Symmetry - Quantum Boundary" and from the Perimeter Institute.
Research at Perimeter Institute is supported in part by the Government of Canada through the Department of Innovation, Science and Economic Development Canada and by the Province of Ontario through the Ministry of Colleges and Universities.

Y.Y. is partially supported by Japan Society for the Promotion of Science (No.21K13929). 
Y.Y. is also grateful for the invited professor grant from the ENS de Lyon, which funded his visit to the LP ENSL in June-July 2025.

%%%%%%%%%%%%%%%%%%%%%%%%%%%%%%%%%%%%%%%%%%%%%%%%%%%%%%%%%%%%%%%%%%%%%%%
\appendix
\section{Derivation of the Regularity  conditions}
\label{A:regular}
%%%%%%%%%%%%%%%%%%

We derive here the regularity conditions \eqref{cond_a} and \eqref{cond_b}. 
We first consider the case \eqref{ab_r=0}: 
\begin{align}
    \dot a \sim k r^n,~~\frac{d}{dr} \log b \sim k'r^m\nn~~{\rm for}~~r\to0.
\end{align}
Then, the Einstein tensors \eqref{G_ab} become
\begin{align}\label{G_ab_r}
    G^t{}_t&\sim -k r^{n-2},~G^r{}_r\sim -k r^{n-2}+2k'(r-\frac{k}{n+1}r^{n+1})r^{m-2},\nonumber\\
    G^\theta{}_\theta&\sim
    \frac{k'}{2}(2+\frac{k}{n+1}r^n-3kr^n)r^{m-1}-\frac{1}{2}knr^{n-2}\nonumber\\
    &+(1-\frac{k}{n+1}r^n)(k'mr^{m-1}+k'^2r^{2m}).
\end{align}
Using the Einstein equations, the regularity for the energy density $-\cT{}_t{}^t$ and radial pressure $\cT{}_r{}^r$ thus requires 
\begin{equation}\label{nm}
    n\geq 2,~~m\geq 1.
\end{equation}
Then $G^\theta{}_\theta$ is regular, and so is $\cT{}_\theta{}^\theta$. 
The Weyl tensors are given by 
\begin{align}
C_{\mu\nu\a\b}C^{\mu\nu\a\b}
\sim
&
\frac{1}{3r^4}\bigg{[}\frac{k(n-1)(n-2)}{n+1}r^n
\\
&
+2k'^2(-1+\frac{k}{n+1}r^n)r^{2m+2}
\nn\\
&
+k'(2-2m+\frac{k(3n+2m-2)}{n+1}r^n)r^{m+1} \bigg{]}^2
\,,\nn
\end{align}
which is also regular for \eqref{nm}. 

We next discuss another case where 
\begin{equation}
    a(r)\sim \mathrm{const.}=a_0,~~~~\frac{d}{dr} \log b \sim k'r^m
\end{equation}
for $r\to0$.
The Einstein tensors \eqref{G_ab} become
\begin{align}\label{G_ab_r}
    G^t{}_t&\sim 0,~G^r{}_r\sim 2k'(r-a_0)r^{m-2},\\
    G^\theta{}_\theta&\sim \frac{k'}{2}(2r+a_0)r^{m-2}+(1-\frac{a_0}{r})(k'mr^{m-1}+k'^2r^{2m}).
\nonumber
\end{align}
From the second one, we need 
\begin{equation}\label{m}
    m\geq 2,
\end{equation}
which is sufficient to make the third one regular. The Weyl tensor is 
\begin{align}
    C_{\mu\nu\a\b}C^{\mu\nu\a\b}&\sim
    \frac{1}{3r^6}\left[6a_0-2k'^2(r-a_0)r^{2m+2}\right. \\
    &\left. +k'((2m-5)a_0-2(m-1)r)r^{m+1}\right]^2.\nonumber
\end{align}
No matter what value $m$ has, the term $\frac{a_0^2}{r^6}$ remains, causing a singularity unless $a_0=0$. If $a_0=0$, the condition \eqref{m} reduces to $m\geq 1$, which is sufficient to regularize $G^\mu{}_\nu$ and $C_{\mu\nu\a\b}C^{\mu\nu\a\b}$.

The last case we consider is,  
\begin{equation}
    \dot a(r)\sim k r^n,~~ b(r)={\rm{const}}.,
\end{equation}
for $r\to0$. We have 
\begin{align}
    G^t{}_t=G^r{}_r\sim -kr^{n-2},~~G^\theta{}_\theta \sim-\frac{1}{2}knr^{n-2},\nn\\
    C_{\mu\nu\a\b}C^{\mu\nu\a\b}\sim\frac{k^2(n^2-3n+2)}{3(n+1)^2}r^{2n-4}.
\end{align}
The regularity at $r=0$ demands $n\geq2$, which is consistent with the previous studies on regular metrics 
\cite{1968qtr..conf...87B,Dymnikova:1992ux,Ayon-Beato:1998hmi,Dymnikova:2004zc, Hayward:2005gi}. 
%\cite{Bardeen,Dymnikova_1992,Ayon-Beato,Dymnikova_2004,Hayward_BH}.

Thus, we conclude with the conditions \eqref{cond_a} and \eqref{cond_b}.

%%%%%%%%%%%%%%%%%%%%%%%%%%%%%%%%%%%%%%%%%%%%%%%%%%%%%%%%%%%%%%%%%%%%%%%%%
\stoptoc

\bibliographystyle{bib-style}
\bibliography{Heff-project}

\end{document}